\documentclass[reprint, superscriptaddress, nofootinbib, altaffilsymbol, aps, prc, floatfix]{revtex4-1}
\pdfoutput=1
\usepackage[colorlinks=true,pdfstartview=FitV,linkcolor=blue,citecolor=blue,urlcolor=blue,bookmarks=false]{hyperref}
\usepackage[usenames,dvipsnames]{color}
\usepackage[sc]{mathpazo}
\usepackage[below]{placeins}
\usepackage{afterpage}
\usepackage[separate-uncertainty,retain-explicit-plus, per-mode = symbol, detect-weight=true, range-units=brackets]{siunitx}
\usepackage{multirow}
\usepackage{rotating}
\usepackage[displaymath, mathlines]{lineno}
\usepackage{graphicx}
\usepackage{paralist}
\usepackage{wrapfig}
\usepackage{appendix}
\usepackage{vruler}
\usepackage{etoolbox}
\usepackage{tikz}
\usepackage{listings}
\usepackage{amsmath}
\usepackage{relsize}
\usepackage{etoolbox}
\usepackage{makecell}
\usepackage{gensymb}

\modulolinenumbers[5]
\newcommand{\trit}{$^3$H}
\newcommand{\ber}{$^7$Be}
\newcommand{\sod}{$^{22}$Na}

\newcommand{\gray}{$\gamma$-ray}

\DeclareSIUnit\kgsi{kg_\text{Si}}
\DeclareSIUnit\bqkg{Bq/\kgsi}
\DeclareSIUnit\atom{atom}
\DeclareSIUnit\atoms{atoms}
\DeclareSIUnit\atomstrit{atoms (^{3}H)}
\DeclareSIUnit\atomsber{atoms (^{7}Be)}
\DeclareSIUnit\atomssod{atoms (^{22}Na)}
\DeclareSIUnit\neutrons{neutrons}
\DeclareSIUnit\muons{\micro^-}
\DeclareSIUnit\gpg{g/g}
\DeclareSIUnit\c{$c$}
\DeclareSIUnit\second{sec}
\DeclareSIUnit\day{day}
\DeclareSIUnit\dayshort{d}
\DeclareSIUnit\week{w}
\DeclareSIUnit\year{yr}
\DeclareSIUnit\standard{std}
\DeclareSIUnit\str{sr}
\DeclareSIUnit\eqquanta{eq.q.}
\DeclareSIUnit\decays{decays}

\newif\ifcolorfigs
\colorfigstrue        %Online Version
\begin{document}

\title{Cosmogenic activation of silicon}
%---

%%%%%%%%%%%%%%%%%%%%%%%%%%%%%%%%%%%%%%%%%%%%%%%%%%%%%%%%%%%%%%%%%%%%%%%%%
%--- Version for revtex

\newcommand{\pnnl}{Pacific Northwest National Laboratory, Richland, Washington 99352, USA}
\newcommand{\lanl}{Los Alamos National Laboratory, Los Alamos, New Mexico 87545, USA}
\newcommand{\uw}{Center for Experimental Nuclear Physics and Astrophysics, University of Washington, Seattle, Washington 98195, USA}
\newcommand{\uc}{Kavli Institute for Cosmological Physics and The Enrico Fermi Institute, The University of Chicago, Chicago, Illinois 60637, USA}
\newcommand{\ua}{Department of Physics and Astronomy, University of Alabama, Tuscaloosa, Alabama 35487, USA}
\newcommand{\disney}{Duckburg, Calisota}
\author{R. Saldanha}\email[Corresponding author: ]{richard.saldanha@pnnl.gov}\affiliation{\pnnl}
\author{R. Thomas}\affiliation{\uc}
\author{R.H.M. Tsang} \altaffiliation[Now at: ]{\ua}\affiliation{\pnnl}
\author{A.E. Chavarria}\affiliation{\uw}
\author{R. Bunker}\affiliation{\pnnl}
\author{J. L. Burnett}\affiliation{\pnnl}
\author{S.R.~Elliott}\affiliation{\lanl}
\author{A. Matalon}\affiliation{\uc}
\author{P. Mitra}\affiliation{\uw}
\author{A. Piers}\affiliation{\uw}
\author{P. Privitera}\affiliation{\uc}
\author{K. Ramanathan}\affiliation{\uc}
\author{R. Smida}\affiliation{\uc}

%%%%%%%%%%%%%%%%%%%%%%%%%%%%%%%%%%%%%%%%%%%%%%%%%%%%%%%%%%%%%%%%%%%%%%%%%

\begin{abstract}
The production of $^{3}$H, $^{7}$Be, and $^{22}$Na by interactions of cosmic-ray particles with silicon can produce radioactive backgrounds in detectors used to search for rare events. Through controlled irradiation of silicon CCDs and wafers with a neutron beam that mimics the cosmic-ray neutron spectrum, followed by direct counting, we determined that the production rate from cosmic-ray neutrons at sea level is ($112 \pm 24$) atoms/(kg day) for $^{3}$H, ($8.1 \pm 1.9 $) atoms/(kg day) for $^{7}$Be, and ($43.0 \pm 7.2 $) atoms/(kg day) for $^{22}$Na. Complementing these results with the current best estimates of activation cross sections for cosmic-ray particles other than neutrons, we obtain a total sea-level cosmic-ray production rate of ($124 \pm 25$) atoms/(kg day) for $^{3}$H, ($9.4 \pm 2.0$) atoms/(kg day) for $^{7}$Be, and ($49.6 \pm 7.4$) atoms/(kg day) for $^{22}$Na.
These measurements will help constrain background estimates and determine the maximum time that silicon-based detectors can remain unshielded during detector fabrication before cosmogenic backgrounds impact the sensitivity of next-generation rare-event searches.
\end{abstract}
%---
%---


%---
\maketitle

\section{Introduction}
\label{sec:intro}
Interactions of cosmic-ray particles with detector materials can produce radioactive isotopes that create backgrounds for experiments searching for rare events such as dark matter interactions and neutrinoless double beta decay. Silicon is a widely used detector material because it is available with very high purity, which leads to low intrinsic radioactive backgrounds. In particular, solid-state silicon-based detector technologies show promise because their eV-scale energy thresholds~\cite{PhysRevLett.123.181802,Abramoff:2019dfb,Agnese:2018col} provide sensitivity to scattering events between atoms and ``low-mass'' dark matter particles with masses below 1\,GeV/c$^{2}$~\cite{Essig:2015cda}.

Three prominent low-mass dark matter efforts that employ silicon detectors are DAMIC~\cite{aguilararevalo2020results}, SENSEI~\cite{Abramoff:2019dfb}, and SuperCDMS~\cite{PhysRevD.95.082002}. All three use the highest-purity single-crystal silicon as detector substrates~\cite{VONAMMON198494}, with sensors fabricated on the surfaces for readout of charge or phonons and installed in low-background facilities to reduce the event rate from environmental backgrounds.

 A primary challenge in these rare-event searches is to distinguish potential signal events from the much higher rate of interactions due to conventional sources of radiation, both from the terrestrial environment and in the detector materials. A variety of mitigation strategies are used to minimize backgrounds; nevertheless, a nonzero residual background expectation is generally unavoidable. Beta-emitting radiocontaminants in the bulk and on the surfaces of the detectors are especially challenging in the search for dark matter because the decay products can produce energy signals that are indistinguishable from the expected signal. Both DAMIC and SuperCDMS have investigated these detector backgrounds (see, e.g., Refs.~\cite{Aguilar-Arevalo:2015lvd,aguilararevalo2020results,PhysRevD.95.082002,Orrell:2017rid}), and they have identified \trit~(tritium), $^{32}$Si (intrinsic to the silicon) and $^{210}$Pb (surface contamination) as the leading sources of background for future silicon-based dark matter experiments. Unlike for $^{32}$Si, there are not yet any direct measurements of the tritium background in silicon; current estimates are based on models that have yet to be validated.
 
Tritium and other radioactive isotopes such as \ber~and \sod~are produced in silicon detectors as a result of cosmic-ray exposure, primarily due to interactions of high-energy cosmic-ray neutrons with silicon nuclei in the detector substrates~\cite{cebrian,Agnese:2018kta}.  %It is expected that effectively all tritium atoms present in the raw silicon material are driven out during single-crystal growth. Consequently, 
The level of background from cosmogenic isotopes in the final detector is effectively determined by the above-ground exposure time during and following detector production, the cosmic-ray flux, and the isotope-production cross sections.  The neutron-induced production cross sections for tritium, \ber, and to a lesser extent \sod, are not experimentally known except for a few measurements at specific energies.  There are several estimates of the expected cross sections; however, they vary significantly, leading to large uncertainties in the expected cosmogenic background for rare-event searches that employ silicon detectors. To address this deficiency, we present measurements of the integrated isotope-production rates from a neutron beam at the Los Alamos Neutron Science Center (LANSCE) ICE HOUSE facility \cite{lisowski2006alamos, icehouse}, which has a similar energy spectrum to that of cosmic-ray neutrons at sea level. This spectral-shape similarity allows for a fairly direct extrapolation from the measured beam production rates to the expected cosmogenic production rates. While the spectral shape is similar, the flux of neutrons from the LANSCE beam greater than \SI{10}{MeV} is roughly \num{5E8} times larger than the cosmic-ray flux, which enables production of measurable amounts of cosmogenic isotopes in short periods of time. Our measurement will allow the determination of acceptable above-ground residency times for future silicon detectors, as well as improve cosmogenic-related background estimates and thus sensitivity forecasts. 

We begin in Sec.~\ref{sec:isotopes} with a discussion of radioisotopes that can be cosmogenically produced in silicon, and we identify those most relevant for silicon-based dark matter searches:  \trit, \ber, and \sod.  For these three isotopes, we review previous measurements of the production cross sections and present the cross-section models that we use in our analysis.  Section~\ref{sec:exposure} introduces our experimental approach, in which several silicon targets---a combination of charge-coupled devices (CCDs) and wafers---were irradiated at LANSCE. In Sec.~\ref{sec:counting} and Sec.~\ref{sec:production_rates} we present our measurements and predictions of the beam-activated activities, respectively. These results are combined in Sec.~\ref{sec:cosmogenic_rates} to provide our best estimates of the production rates from cosmogenic neutrons. In Sec.~\ref{sec:alternate} we evaluate other (non-neutron)  production mechanisms and we conclude in Sec.~\ref{sec:discussion} with a summarizing discussion.

\section{Cosmogenic Radioisotopes}
\label{sec:isotopes}

\begin{table}[t]
\centering
\begin{tabular}{c c c c}
\hline
Isotope & Half-life & Decay & Q-value \\
 & [yrs] & mode & [keV]\\
\hline
\vrule width 0pt height 2.2ex
%$^3$H & \num{12.32 \pm 0.02} & $\beta$- & \num{18.591 \pm 0.003} \\
%$^7$Be & \num{0.1457 \pm 0.002} & EC & \num{861.82 \pm 0.02}\\
%$^{10}$Be & \num{1.51 \pm 0.06 E6} &$\beta$- & \num{556.0 \pm 0.6}\\
%$^{14}$C & \num{5700 \pm 30} & $\beta$- & \num{156.475 \pm 0.004}\\
%$^{22}$Na & \num{2.6018 \pm 0.0022} & $\beta$+ & \num{2842.2 \pm 0.2}\\
%$^{26}$Al & \num{7.17 \pm 0.24 E5} & EC & \num{4004.14 \pm 6}\\
$^3$H & 12.32\,$\pm$\,0.02 & $\beta$- & 18.591\,$\pm$\,0.003 \\
$^7$Be & 0.1457\,$\pm$\,0.0020 & EC & 861.82\,$\pm$\,0.02\\
$^{10}$Be & (1.51\,$\pm$\,0.06)$\times$10$^6$ & $\beta$- & 556.0\,$\pm$\,0.6\\
$^{14}$C & 5700\,$\pm$\,30 & $\beta$- & 156.475\,$\pm$\,0.004\\
$^{22}$Na & 2.6018\,$\pm$\,0.0022 & $\beta$+ & 2842.2\,$\pm$\,0.2\\
$^{26}$Al & (7.17\,$\pm$\,0.24)$\times$10$^5$ & EC & 4004.14\,$\pm$\,6.00\\
%$^{32}$Si & \num{153 \pm 19} & $\beta$- & \num{227.2 \pm 0.3}\\
%$^{32}$P & \num{3.906 \pm 0.001 E-2} & $\beta$- & \num{1710.66 \pm 0.04}\\
\hline
\end{tabular}
\caption{List of all radioisotopes with half-lives $>$\,30 days that can be produced by cosmogenic interactions with natural silicon. All data is taken from NNDC databases \cite{dunford1998online}. \protect\footnotemark[1]}
\footnotetext{Unless stated otherwise, all uncertainties quoted in this paper are at 1$\sigma$ (68.3\%) confidence.}
\label{tab:rad_isotopes}
\end{table}

Most silicon-based dark matter experiments use high-purity ($\gg$\,99\%) natural silicon (92.2\% $^{28}$Si, 4.7\% $^{29}$Si, 3.1\% $^{30}$Si \cite{meija2016isotopic}) as the target detector material. The cosmogenic isotopes of interest for these experiments are therefore any long-lived radioisotopes that can be produced by cosmic-ray interactions with silicon; Table~\ref{tab:rad_isotopes} lists all isotopes with half-lives greater than 30 days that are lighter than $^{30}$Si + n/p. None of them have radioactive daughters that may contribute additional backgrounds. Assuming that effectively all non-silicon atoms present in the raw material are driven out during growth of the single-crystal silicon boules used to fabricate detectors, and that the time between crystal growth and moving the detectors deep underground is typically less than 10 years, cosmogenic isotopes with half-lives greater than 100 years (i.e., $^{10}$Be, $^{14}$C, and $^{26}$Al) do not build up sufficient activity~\cite{reedy2013cosmogenic, caffee2013cross} to produce significant backgrounds. Thus the cosmogenic isotopes most relevant to silicon-based rare-event searches are tritium, \ber, and \sod. Tritium is a particularly dangerous background for dark matter searches because it decays by pure beta emission and its low Q-value (\SI{18.6} {\keV}) results in a large fraction of decays that produce low-energy events in the expected dark matter signal region. \ber~decays by electron capture, either directly to the ground state of $^7$Li (89.56\%) or via the \SI{477}{\keV} excited state of $^7$Li (10.44\%). \ber~is not a critical background for dark matter searches, because it has a relatively short half-life (\SI{53.22}{\day}); however, the \SI{54.7}{\eV} atomic de-excitation following electron capture may provide a useful energy-calibration tool. \sod~decays primarily by positron emission (90.3\%) or electron capture (9.6\%) to the 1275 keV level of $^{22}$Ne. For thin silicon detectors \sod~can be a significant background as it is likely that both the \SI{1275}{\keV} $\gamma$ ray and the \SI{511}{\keV} positron-annihilation photons will escape undetected, with only the emitted positron or atomic de-excitation following electron capture depositing any energy in the detector. Note that compared to \trit, the higher $\beta^+$ endpoint (\SI{546}{keV}) means that a smaller fraction of the \sod~decays produce signals in the energy range of interest for dark matter searches. 

\subsection{Tritium Production}
\begin{figure}[t!]
   \centering
   \includegraphics[width=\columnwidth]{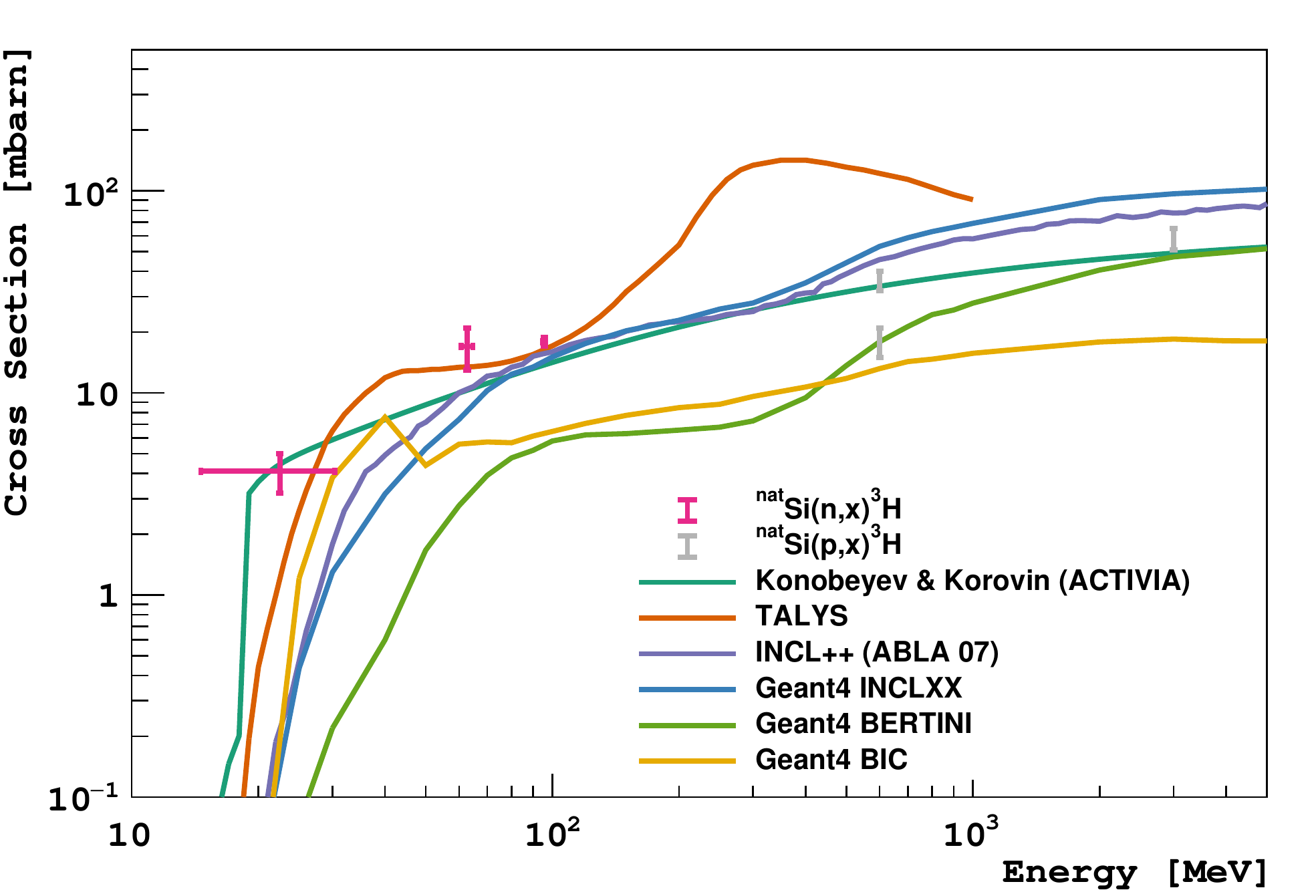} % requires the graphicx package
   %\captionsetup{width=0.9\textwidth}
   \caption{Experimental measurements (magenta error bars) \cite{QAIM1978150, Tippawan:2004sy, benck2002secondary} and model estimates (continuous curves) of neutron-induced tritium production in silicon. Measurements of the proton-induced cross section \cite{goebel1964production, kruger1973high} are also shown for reference (gray error bars).}
   \label{fig:si_3h_cross_sections}
\end{figure}

Tritium production in silicon at sea-level is dominated by spallation interactions of high-energy cosmogenic neutrons with silicon nuclei. Tritium is a pure $\beta$ emitter and it is therefore not possible to directly measure the production cross section using conventional methods that rely on \gray~detectors to tag the reaction products. There are three previous experimental measurements of the neutron-induced tritium production cross section in silicon (shown in Fig.~\ref{fig:si_3h_cross_sections}), which either extracted tritium from a silicon target and measured the activity in a proportional counter \cite{QAIM1978150} or measured the triton nuclei ejected from a silicon target using $\Delta E-E$ telescopes \cite{Tippawan:2004sy,benck2002secondary}. The proton-induced cross section is expected to be similar to that of neutrons so we also show previous measurements with proton beams \cite{goebel1964production, kruger1973high}. While these measurements provide useful benchmarks at specific energies, they are insufficient to constrain the cosmogenic production cross section across the full range of relevant neutron energies (from $\sim$10\,MeV to a few GeV).

For this reason, previous estimates of tritium production in silicon dark matter detectors have relied on estimates of the cross section from calculations and simulations of the nuclear interactions or compiled databases that combine calculations with experimental data \cite{martoff1987limits, zhang2016cosmogenic, agnese2019production}. The production of tritons due to spallation is difficult to model, because the triton is a very light nucleus that is produced not only during the evaporation or de-excitation phase but also from coalescence of nucleons emitted during the high-energy intra-nuclear cascade stage \cite{leray2010improved, leray2011results, filges2009handbook}. Due to large variations among the predictions of different cross-section models, we consider several models for comparison to our experimental results and extraction of cosmogenic production rates. Shown in Fig.~\ref{fig:si_3h_cross_sections} are the semi-empirical formulae of Konobeyev and Korovin (K\&K) \cite{konobeyev1993tritium} (extracted from the commonly used ACTIVIA code \cite{back2008activia}) and results from nuclear reaction calculations and Monte Carlo simulations that are performed by codes such as TALYS \cite{koning2008talys}, INCL \cite{boudard2013new} and ABLA \cite{kelic2008deexcitation}.\footnote{The Konobeyev and Korovin (\trit), and Silberberg and Tsao (\ber, \sod) cross sections were obtained from the ACTIVIA code package \cite{activia2017}, the TALYS cross sections were calculated using TALYS-1.9 \cite{talys1.9}, and the INCL cross sections were calculated using the INCL++ code (v6.0.1) with the ABLA07 de-excitation model \cite{mancusi2014extension}. The default parameters were used for all programs. We note that the TALYS models are optimized in the \SI{1}{\keV} to \SI{200}{\MeV} energy range though the maximum energy has been formally extended to \SI{1}{\GeV} \cite{koning2014extension}.} We also compared effective cross sections (extracted through simulation) from built-in physics libraries of the widely used Geant4 simulation package \cite{agostinelli2003geant4,allison2016recent} such as INCLXX \cite{boudard2013new,mancusi2014extension}, BERTINI \cite{bertini1963low, guthrie1968calculation, bertini1969intranuclear, bertini1971news}, and Binary Cascades (BIC) \cite{folger2004binary}.\footnote{We used Geant4.10.3.p02 with physics lists QGSP\_INCLXX 1.0 (INCL++ v5.3), QGSP\_BERT 4.0, and QGSP\_BIC 4.0.}

\subsection{\ber~Production}
\begin{figure}[t]
   \centering
   \includegraphics[width=\columnwidth]{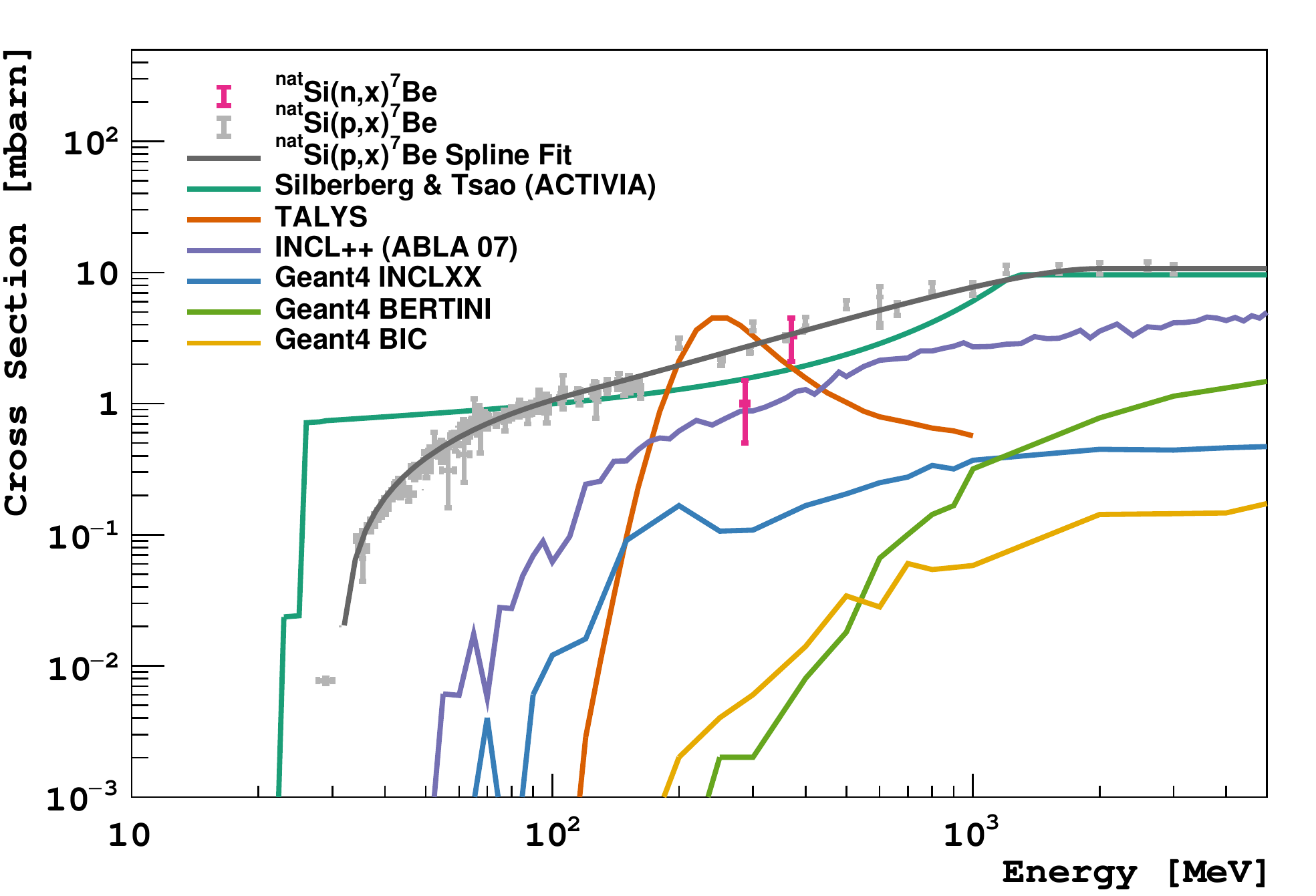} 
   \caption{Experimental measurements (magenta error bars) \cite{ninomiya2011cross} and model estimates (continuous curves) of the neutron-induced \ber~production cross section in silicon. Measurements of the proton-induced cross section \cite{otuka2014towards, zerkin2018experimental} are also shown for reference (gray error bars).}
   \label{fig:si_7be_cross_sections}
\end{figure}

\ber~is produced as an intermediate-mass nuclear product of cosmogenic particle interactions with silicon. The neutron-induced production cross section has been measured at only two energies \cite{ninomiya2011cross}, as shown in Fig.~\ref{fig:si_7be_cross_sections}. Although the neutron- and proton-induced cross sections are not necessarily the same, especially for neutron-deficient nuclides such as \ber~and \sod~\cite{ninomiya2011cross}, there are a large number of measurements with protons 
%of the proton-induced production cross-section 
that span the entire energy range of interest \cite{otuka2014towards, zerkin2018experimental}, which we show in Fig.~\ref{fig:si_7be_cross_sections} for comparison.\footnote{We have excluded measurements from Ref.~\cite{rayudu1968formation}, because there are well-known discrepancies with other measurements \cite{ michel1995nuclide, schiekel1996nuclide}.} For ease of evaluation, we fit the proton cross-section data with a continuous 4-node spline, hereafter referred to as ``$^{\text{nat}}$Si(p,x)$^7$Be Spline Fit''. 
As with tritium, we also show predictions from different nuclear codes and semi-empirical calculations, including the well-known Silberberg and Tsao (S\&T) semi-empirical equations \cite{silberberg1973partial,silberberg1973partial2, silberberg1977cross, silberberg1985improved, silberberg1990spallation, silberberg1998updated} as implemented in the ACTIVIA code. We note that the model predictions for the \ber~production cross section in silicon vary greatly, with significantly different energy thresholds, energy dependence, and magnitude. \ber~is believed to be produced predominantly as a fragmentation product rather than as an evaporation product or residual nucleus \cite{michel1995nuclide}, and fragmentation is typically underestimated in most theoretical models \cite{michel1995nuclide, titarenko2006excitation}. We note that unlike for the tritium cross-section models, there is a significant difference between the predictions obtained by evaluating the INCL++ v6.0.1 model directly versus simulating with Geant4 (INCL++ v5.3), probably due to updates to the model.

\subsection{\sod~Production}
\begin{figure}[t]
   \centering
   \includegraphics[width=\columnwidth]{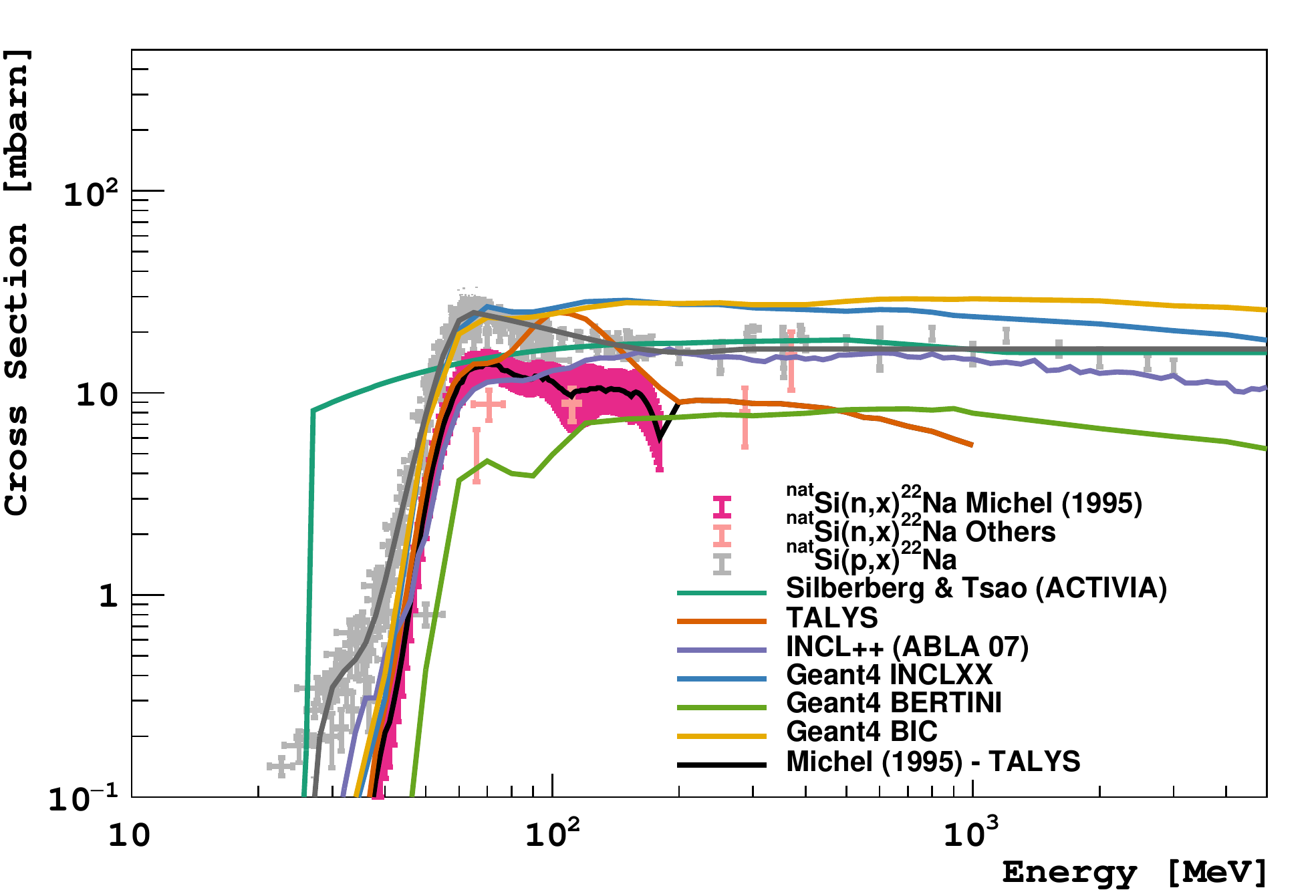} % requires the graphicx package
   %\captionsetup{width=0.9\textwidth}
   \caption{Experimental measurements (magenta and pink error bars) \cite{michel2015excitation, hansmann2010production, yashima2004measurement, sisterson2007cross, ninomiya2011cross} and model estimates (continuous curves) of the neutron-induced \sod~production cross section in silicon. Measurements of the proton-induced cross section \cite{otuka2014towards, zerkin2018experimental} are also shown for reference (gray error bars).}
   \label{fig:si_22na_cross_sections}
\end{figure}

\sod~is produced as a residual nucleus following cosmogenic interactions with silicon. Compared to tritium and \ber, the production of \sod~is the best studied. Measurements of the neutron-induced cross section were carried out by Michel et.\ al.\ using quasi-monoenergetic neutrons between 33 and 175 MeV, with TALYS-predicted cross sections used as the initial guess to unfold the experimentally measured production yields \cite{michel2015excitation, hansmann2010production}. These, along with six other data points between 66 and 370 MeV \cite{yashima2004measurement, sisterson2007cross, ninomiya2011cross}, are shown in Fig.~\ref{fig:si_22na_cross_sections}. Proton-induced cross-section measurements\footnote{Similar to \ber, we have excluded measurements from Ref.~\cite{rayudu1968formation}.} \cite{otuka2014towards, zerkin2018experimental} span the entire energy range of interest and are significantly larger than the measured neutron-induced cross sections. As before, we also show the predicted cross sections from Silberberg and Tsao, TALYS, INCL++ (ABLA07) and Geant4 models. In order to compare the existing neutron cross-section measurements to our data, we use a piecewise model that follows the measurements in Refs.~\cite{michel2015excitation, hansmann2010production} below 180\,MeV and follows the TALYS model at higher energies. This model is hereafter referred to as ``Michel-TALYS'' (see  Fig.~\ref{fig:si_22na_cross_sections}). \sod~can also be produced indirectly through the production of the short-lived isotopes $^{22}$Mg, $^{22}$Al, and $^{22}$Si, which eventually decay to \sod, but for the models considered the total contribution from these isotopes is $<$ \SI{1}{\percent}, and is ignored here.

\section{Beam Exposure}
\label{sec:exposure}
To evaluate the production rate of cosmogenic isotopes through the interaction of high-energy neutrons, we irradiated silicon charge-coupled devices (CCDs) and silicon wafers at the LANSCE neutron beam facility. Following the irradiation, the CCDs were readout to measure the beam-induced $\beta$ activity within the CCD active region, and the $\gamma$ activity induced in the wafers was measured using $\gamma$-ray spectroscopy. In this section we describe the details of the targets and beam exposure, while in Sec.~\ref{sec:counting} we present the measurement results.

%Silicon charge-coupled devices (CCDs) and silicon wafers were irradiated at the LANSCE neutron beam facility to evaluate the production rate of isotopes through the interactions of fast neutrons.
\subsection{CCDs}
\label{sec:ccds}

The irradiated CCDs were designed and procured by Lawrence Berkeley National Laboratory (LBNL)~\cite{ccdtech} for the DAMIC Collaboration.
CCDs from the same fabrication lot were extensively characterized in the laboratory and deployed underground at SNOLAB to search for dark matter~\cite{Aguilar-Arevalo:2016zop, PhysRevD.94.082006}.
The devices are three-phase scientific CCDs with a buried $p$-channel fabricated on a \SI{670}{\micro\meter}-thick $n$-type high-resistivity (10--20\,\si{\kilo\ohm\cm}) silicon substrate, which can be fully depleted by applying $>$\,\SI{40}{\volt} to a thin backside contact.
The CCDs feature a 61.44$\times$30.72\,mm$^2$ rectangular array of 4096$\times$2048 pixels (each 15$\times$15 \si{\micro\meter\squared}) and an active thickness of \SI{661 \pm 10}{\micro\meter}.
By mass, the devices are $>$\,\SI{99}{\percent} elemental silicon with natural isotopic abundances. Other elements present are oxygen ($\sim$\,\SI{0.1}{\percent}) and nitrogen ($<$\,\SI{0.1}{\percent}) in the dielectrics, followed by phosphorous and boron dopants ($<$\,\SI{0.01}{\percent}) in the silicon.

Ionizing particles produce charge in the CCD active region; e.g., a fast electron or $\beta$ particle will produce on average one electron-hole pair for every \SI{3.8}{\eV} of deposited energy. The ionization charge is drifted by the applied electric field and collected on the pixel array. The CCDs are read out serially by moving the charge vertically row-by-row into the serial register (the bottom row) where the charge is moved horizontally pixel-by-pixel to the output readout node.
Before irradiation, the charge-transfer inefficiency from pixel to pixel was $< 10^{-6}$~\cite{ccdtech}, the dark current was $<$\SI{1}{e^- \per pixel \per \hour}, and the uncertainty in the measurement of the charge collected by a pixel was $\sim$2\,$e^-$ RMS. Further details on the response of DAMIC CCDs can be found in Sec.~IV of Ref.~\cite{PhysRevD.94.082006}.
Even after the significant increase in CCD noise following irradiation (e.g., due to shot noise associated with an increase in dark current), the CCD can still resolve most of the tritium $\beta$-decay spectrum.  

Irradiation generates defects in silicon devices that can trap charges and negatively impact the performance of CCDs. Fully depleted devices are resilient to irradiation damage in the bulk silicon because the ionization charge is collected over a short period of time, which minimizes the probability of charge being trapped by defects before it is collected.
For this reason LBNL CCDs have been considered for space-based imaging where the devices are subjected to high levels of cosmic radiation~\cite{snap}.
Measurements at the LBNL cyclotron demonstrated the remarkable radiation tolerance of the CCDs proposed for the SNAP satellite, which follow the same design principles and fabrication process as the DAMIC CCDs.
For the measurements presented in this paper, there is a trade-off between activation rate and CCD performance. 
Higher irradiation leads to a higher activity of radioisotopes in the CCD and hence a lower statistical uncertainty in the measurement.
On the other hand, higher irradiation also decreases the CCD performance, which needs to be modeled and can thus introduce significant systematic uncertainty.

The two most relevant performance parameters affected by the irradiation are the charge-transfer inefficiency (CTI) and the pixel dark current (DC).
Ref.~\cite{snap} provides measurements of CTI and DC after irradiation with 12.5 and \SI{55}{MeV} protons.
Following irradiation doses roughly equivalent to a LANSCE beam fluence of $2.4\times10^{12}$ neutrons above \SI{10}{\MeV}, the CCDs were still functional with the CTI worsened to $\sim$\,$10^{-4}$ and asymptotic DC rates (after days of operation following a room-temperature anneal) increased to $\sim$\SI{100}{e^- \per pixel \per \hour}.
These values depend strongly on the specific CCD design and the operation parameters, most notably the operating temperature.
Considering the available beam time, the range of estimated production rates for the isotopes of interest, and the CCD background rates, we decided to irradiate three CCDs with different levels of exposure, roughly corresponding to $2.4\times10^{12}$, $1.6\times10^{12}$, and $0.8\times10^{12}$ neutrons above \SI{10}{MeV} at the LANSCE neutron beam. Furthermore, we used a collimator (see Sec.~\ref{sec:lansce_beam}) to suppress irradiation of the serial register at the edge of the CCDs by one order of magnitude and thus mitigate CTI in the horizontal readout direction. Following the beam exposure, we found that the least irradiated CCD had an activity sufficiently above the background rate while maintaining good instrumental response and was therefore selected for analysis in Sec.~\ref{sec:ccd_counting}.

The CCDs were packaged at the University of Washington following the procedure developed for the DAMIC experiment.
%The package provides the signals that drive and read the device.
The CCD die and a flex cable were glued onto a silicon support piece such that the electrical contact pads for the signal lines are aligned.
The CCDs were then wedge bonded to the flex cable with \SI{25}{\micro\meter}-thick aluminum wire.
A connector on the tail of the flex cable can be connected to the electronics for device control and readout.

\begin{figure}
\centering
\includegraphics[width=\columnwidth]{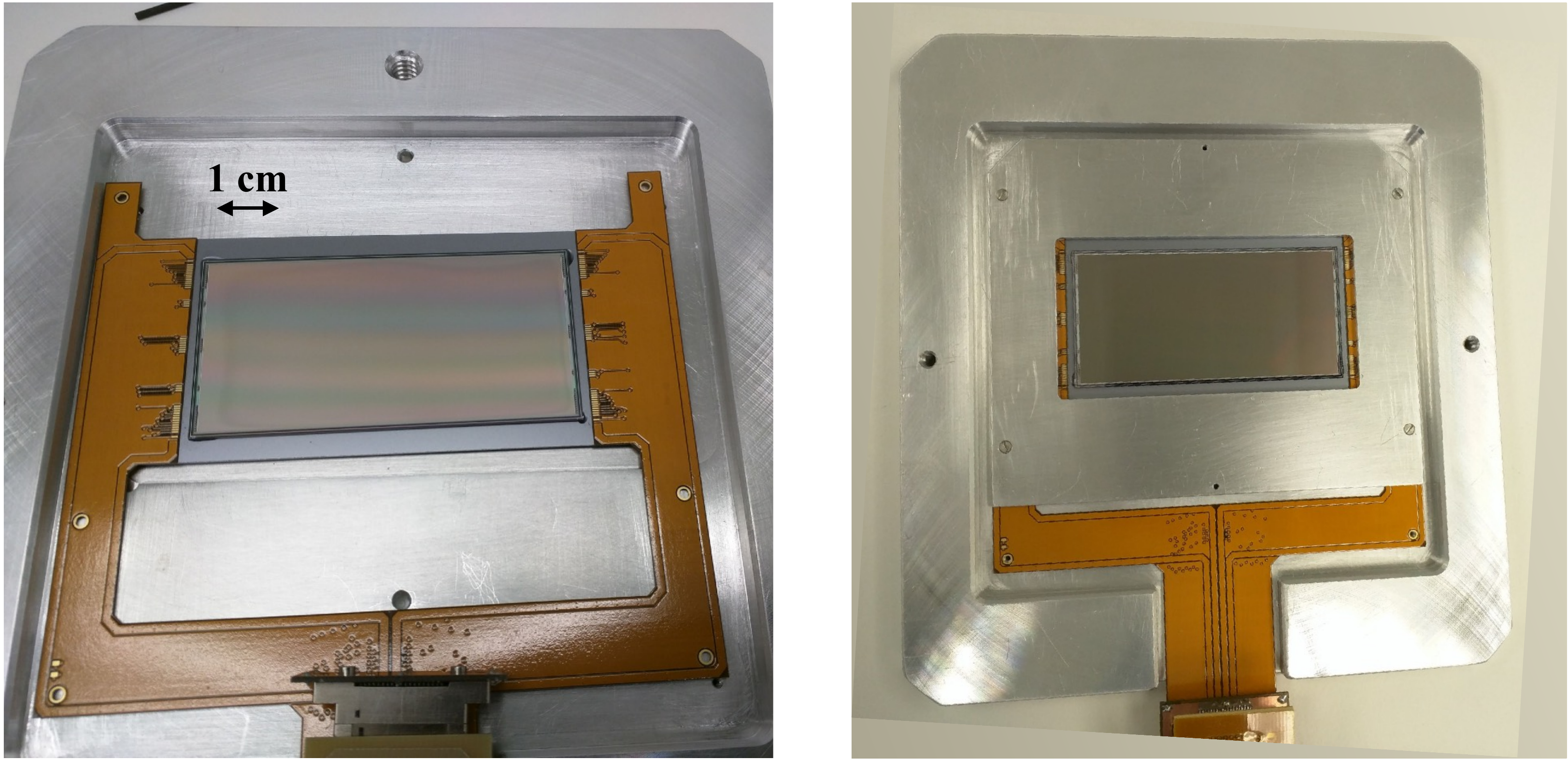}
\caption{Photograph of the CCD package inside its aluminum storage box. Left: Package before wire bonding. Right: After wire bonding, with aluminum frame to keep the CCD package fixed in place.}
\label{fig:CCDphoto}
\end{figure}

Each packaged device was fixed inside an aluminum storage box, as shown in Fig.~\ref{fig:CCDphoto}. The CCDs were kept inside their storage boxes during irradiation to preserve the integrity of the CCD package, in particular to prevent the wire bonds from breaking during handling and to reduce any possibility of electrostatic discharge, which can damage the low-capacitance CCD microelectronics.
To minimize the attenuation of neutrons along the beam path and activation of the storage box, the front and back covers that protect each CCD were made from relatively thin (0.5\,mm) high-purity aluminum (alloy 1100).

\subsection{Wafers}
In addition to the CCDs, we exposed several Si wafers, a Ge wafer, and two Cu plates to the neutron beam. These samples served both as direct targets for activation and measurement of specific radioisotopes, and as witness samples of the neutron beam.  In this paper, we focus on the Si wafers; however, the Ge wafer and Cu plates were also measured and may be the subject of future studies. 

A total of eight Si wafers (4 pairs) were used:  one pair matched to each of the three CCDs (such that they had the same beam exposure time) and a fourth pair that served as a control sample.  The eight wafers were purchased together and have effectively identical properties. Each wafer was sliced from a Czochralski-grown single-crystal boule with a 100-mm diameter and a resistivity of $>$\SI{20}{\ohm\cm}.  The wafers are undoped, were polished on one side, and have a $\langle$100$\rangle$ crystal-plane alignment. The thickness of each individual wafer is \SI{500 \pm 17}{\micro\meter} (based on information from the vendor). The control sample was not exposed to the neutron beam and thus provides a background reference for the gamma counting. Note that because the wafers were deployed and counted in pairs, henceforth we distinguish and refer to only pairs of wafers rather than individual wafers. The (single) Ge wafer is also \SI{100}{\milli\meter} in diameter and undoped, with a thickness of \SI{525 \pm 25}{\micro\meter}, while the Cu plates have dimensions of $114.7 \times 101.6 \times$ \SI{3.175}{\milli\meter}.

\subsection{LANSCE Beam Exposure}
\label{sec:lansce_beam}
\begin{figure*}
\centering
\includegraphics[width=0.32\textwidth]{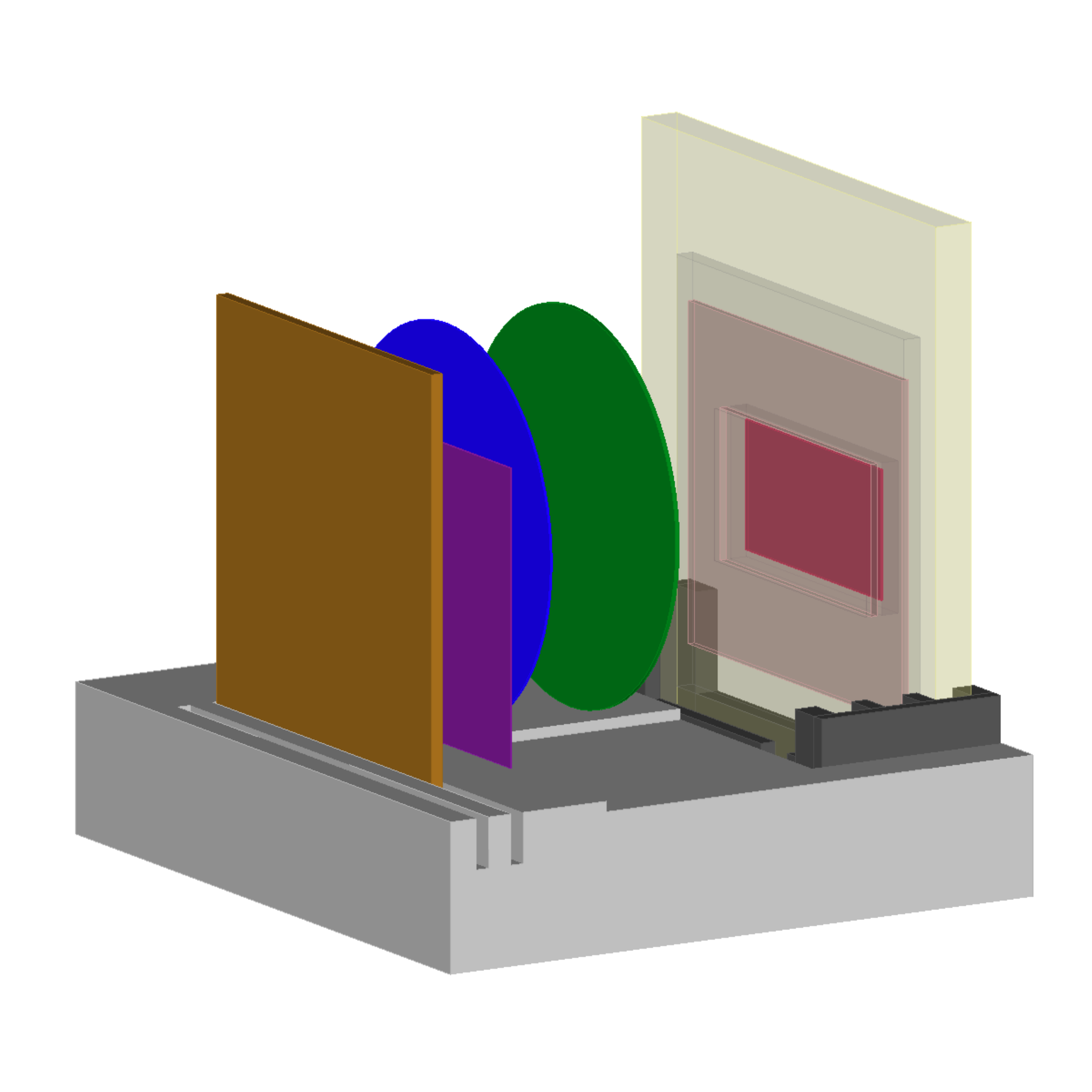}
\includegraphics[width=0.32\textwidth]{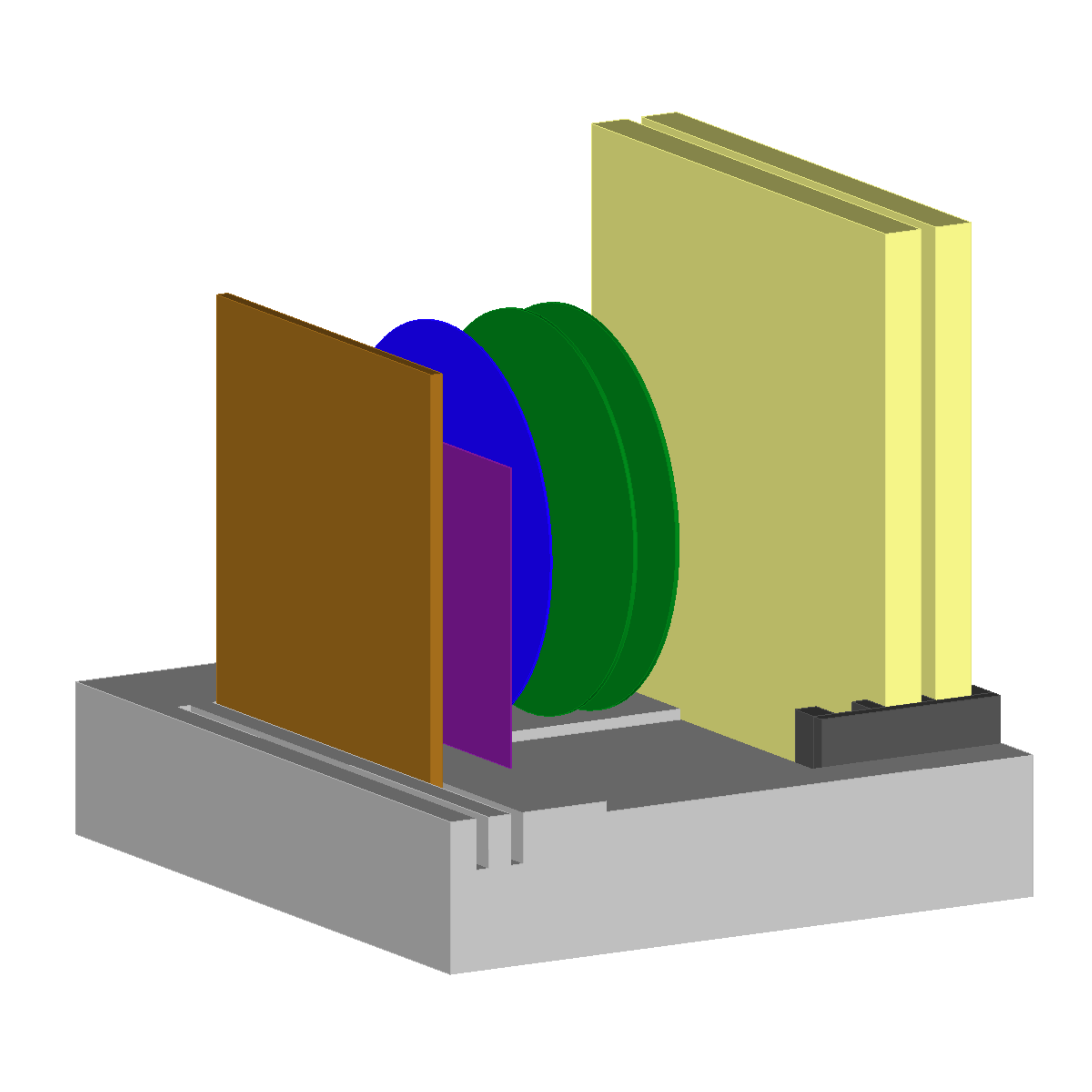}
\includegraphics[width=0.32\textwidth]{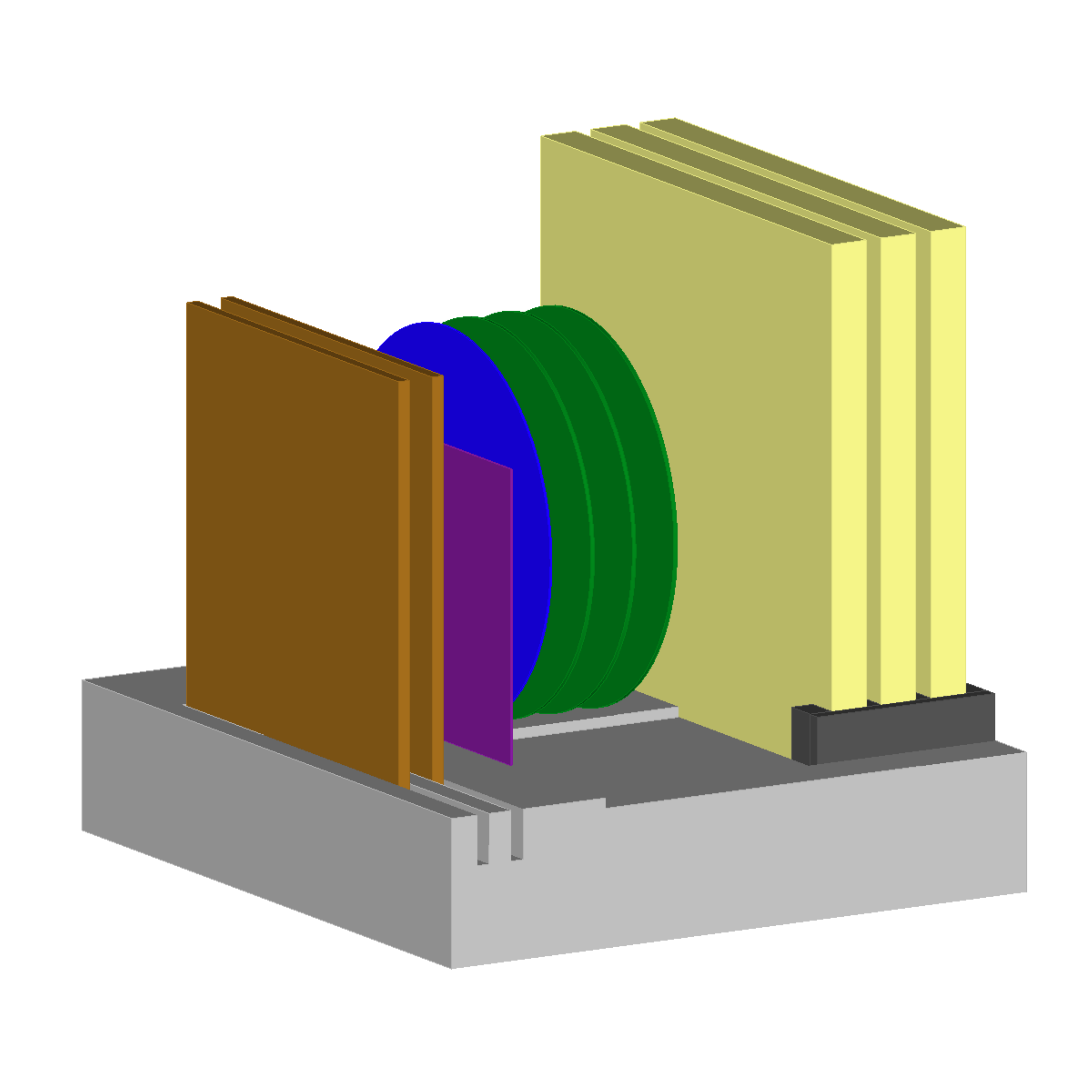}
\caption{Geant4 renderings of the three setups used to position targets in the  neutron beam, with the beam passing from right to left.  
Aluminum (Al) boxes holding the CCDs (yellow) were held in place by an Al rack (dark gray).  
For the initial setup (left), the Al box is made transparent to show the positioning of the CCD (red), air (grey), and other structures (light brown). 
The other targets include pairs of Si wafers (green), a Ge wafer (blue), and Cu plates (copper brown).
The polyethylene wafer holder (purple) is simplified to a rectangle of the same thickness and height as the actual object, with the sides and bottom removed.
All targets were supported on an acetal block (light gray).}
\label{fig:g4rendering}
\end{figure*}
The samples were irradiated at the LANSCE WNR ICE-HOUSE II facility~\cite{icehouse} on Target 4 Flight Path 30 Right (4FP30R). A broad-spectrum (0.2--800 MeV) neutron beam was produced via spallation of 800 MeV protons on a tungsten target. A 2.54-cm (1") diameter beam collimator was used to restrict the majority of the neutrons to within the active region of the CCD and thus prevent unwanted irradiation of the serial registers on the perimeter of the active region. The neutron fluence was measured with $^{238}$U foils by an in-beam fission chamber~\cite{wender1993fission} placed downstream of the collimator. The beam has a pulsed time structure, which allows the incident neutron
energies to be determined using the time-of-flight technique (TOF)---via a measurement between the proton beam pulse and the fission chamber signals~\cite{lisowski2006alamos,wender1993fission}.

\begin{figure}[h!]
\begin{center}
 \includegraphics[width=\columnwidth]{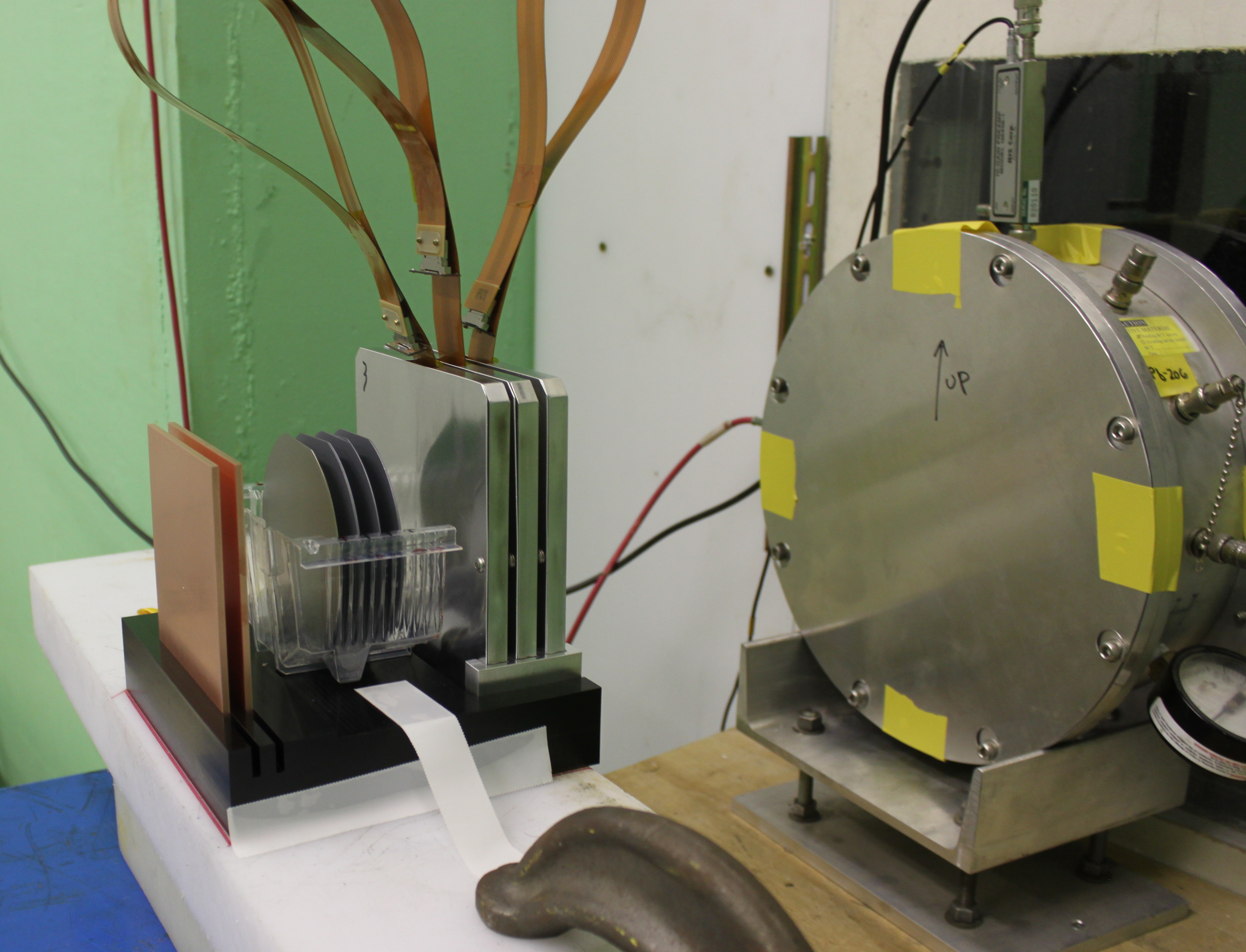}
 \end{center}
 \caption{Layout of the samples as placed in the beam during the final irradiation setup (cf.\ Fig.~\ref{fig:g4rendering} right). The beam first passes through the cylindrical fission chamber (far right) and then through the samples (from right to left): 3~CCDs in Al boxes (with flex cables emerging at the top), 3~pairs of Si wafers, 1~Ge wafer, and 2~Cu plates.}
 \label{Fig:CCDlayout}
\end{figure}

The beam exposure took place over four days between September 18$^{\mathrm{th}}$ and 22$^{\mathrm{nd}}$, 2018. On Sept.\,18, CCD\,1 was placed in the beam line at 18:03 local time, located closest to the fission chamber, along with a pair of Si wafers, one Ge wafer, and one Cu plate placed downstream (in that order; cf.\ Fig.~\ref{fig:g4rendering} left). The front face of the Al box containing CCD\,1 was \SI{260}{\mm} from the face of the fission chamber. At 17:16 on Sept.\,20, CCD\,2 was added directly downstream from CCD\,1, along with another pair of Si wafers. The front face of the Al box for CCD\,2 was \SI{14.3}{\mm} from the front face of CCD\,1. At 09:11 on Sept.\,22, CCD\,3 was added downstream with an equidistant spacing relative to the other CCDs, along with another pair of Si wafers and a second Cu plate. Figure~\ref{fig:g4rendering} shows schematics of these three exposure setups,  while Fig.~\ref{Fig:CCDlayout} shows a photograph of the final setup in which all three CCDs were on the beam line. The exposure was stopped at 08:00 on Sept.\,23, and all parts exposed to the beam were kept in storage for approximately seven weeks to allow short-lived radioactivity to decay prior to shipment for counting.

\subsection{Target Fluence}
\begin{figure}
\centering
\includegraphics[width=\columnwidth]{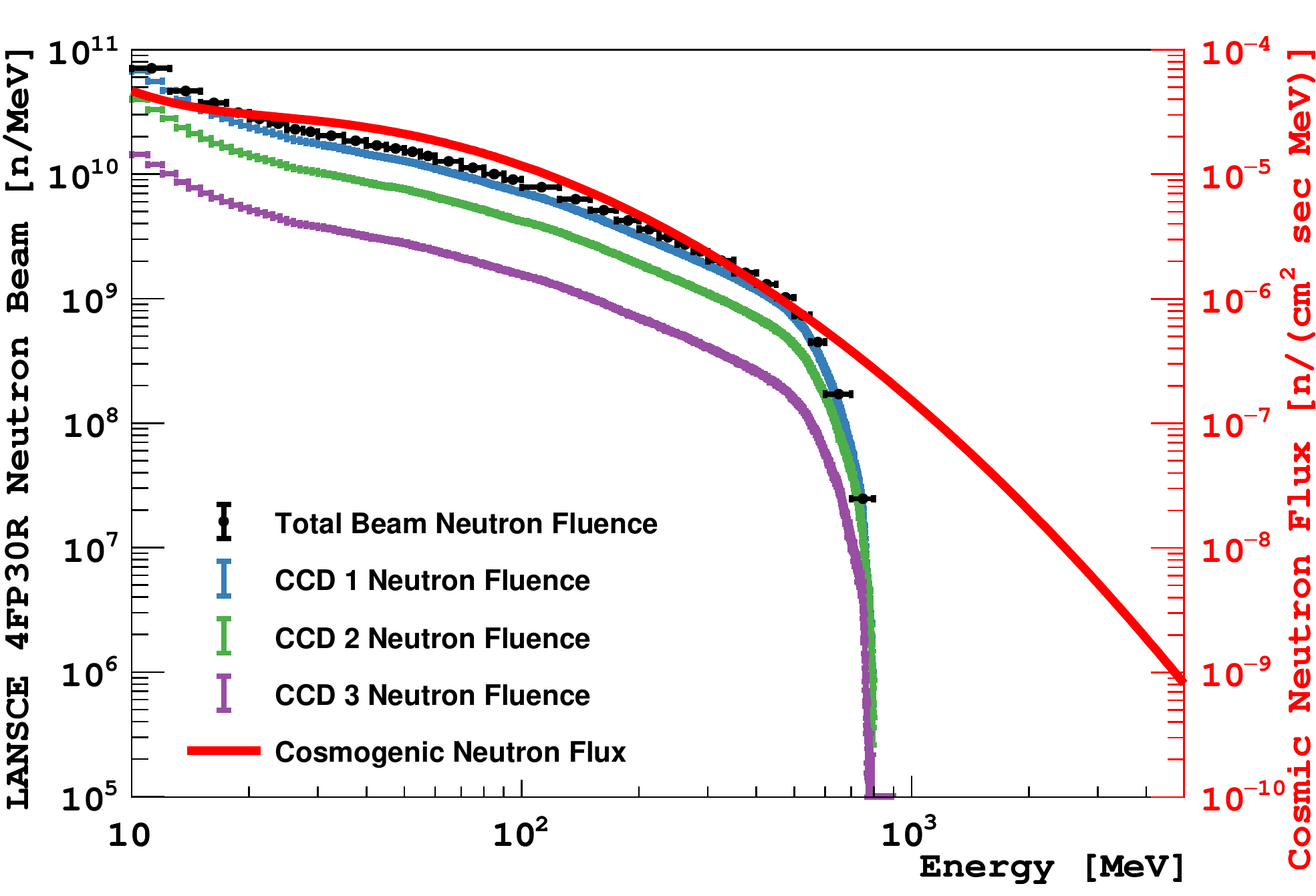}
\caption{Comparison of the LANSCE 4FP30R/ICE II neutron beam with sea-level cosmic-ray neutrons. The black data points and left vertical axis show the number of neutrons measured by the fission chamber during the entire beam exposure used for this measurement. Uncertainties shown are statistical only (see main text for discussion of systematic uncertainties). The colored markers show the simulated fluence for each of the CCDs in the setup. For comparison, the red continuous line and the right vertical axis show the reference cosmic-ray neutron flux at sea level for New York City during the midpoint of solar modulation \cite{gordon2004measurement}}.
\label{fig:lanscebeamenergy}
\end{figure}

\begin{figure}
\begin{center}
 \includegraphics[width=\columnwidth]{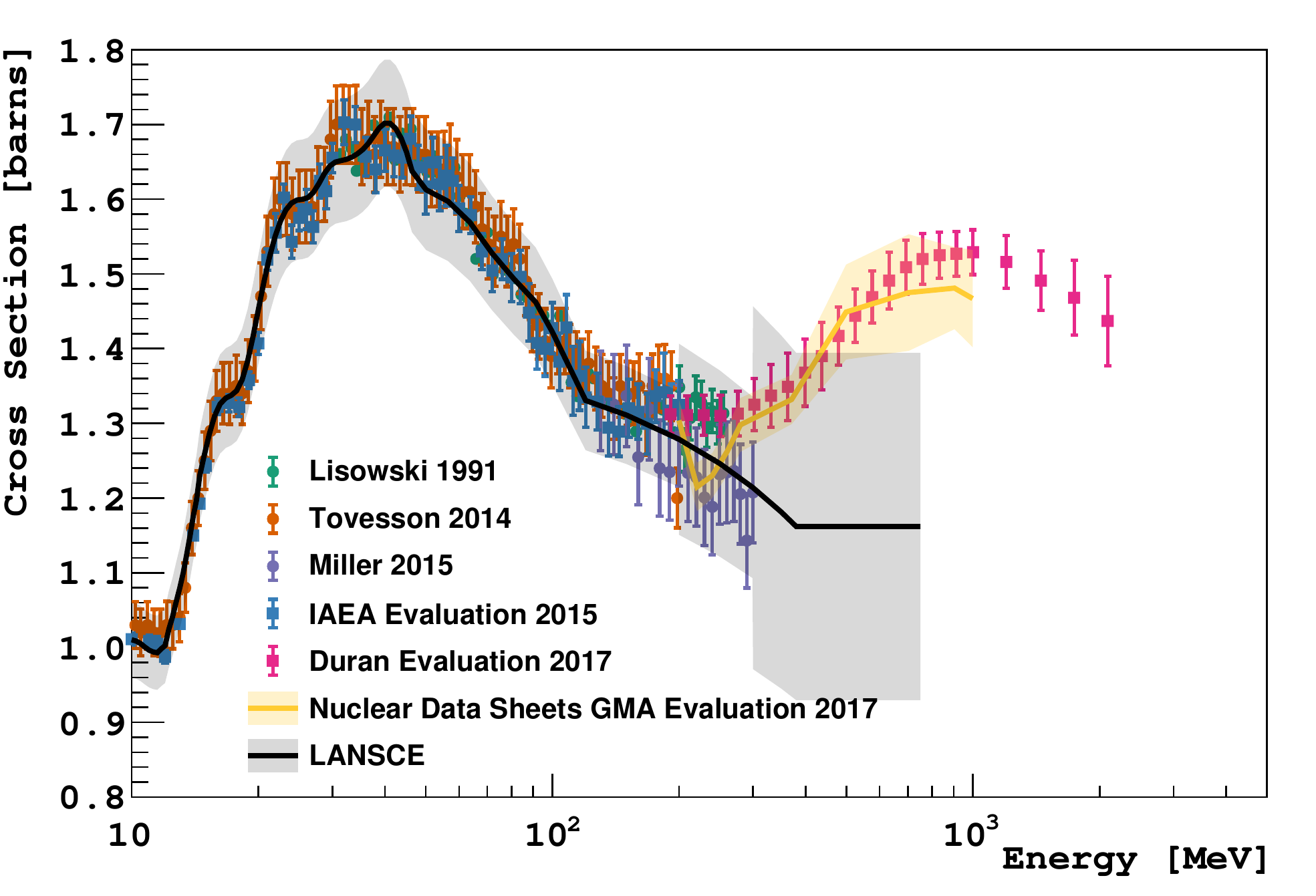}
 \end{center}
 \caption{Experimental measurements (circles) \cite{lisowski1991fission, tovesson2014fast, miller2015measurement} and evaluations (squares) \cite{carlson2009international, marcinkevicius2015209, duran2017search, carlson2018evaluation} of the $^{238}$U(n, f) cross section. The cross section assumed by the LANSCE facility to convert the fission chamber counts to a total neutron fluence is shown by the black line, with the shaded grey band indicating the assumed uncertainty.}
 \label{fig:fission_cs}
\end{figure}

The fluence measured by the fission chamber during the entire beam exposure is shown in Fig.~\ref{fig:lanscebeamenergy}, with a total of \num{2.91 \pm 0.22 E12} neutrons above 10 MeV. The uncertainty is dominated by the systematic uncertainty in the $^{238}$U(n, f) cross section used to monitor the fluence, shown in Fig.~\ref{fig:fission_cs}. Below 200 MeV the assumed LANSCE cross section and various other experimental measurements and evaluations \cite{lisowski1991fission, carlson2009international, tovesson2014fast, marcinkevicius2015209} agree to better than 5\%. Between 200 and 300 MeV there are only two measurements of the cross section \cite{lisowski1991fission, miller2015measurement} which differ by 5--10\%. Above \SI{300}{\MeV} there are no experimental measurements. The cross section used by the LANSCE facility assumes a constant cross section above \SI{380}{\MeV} at roughly the same value as that measured at \SI{300}{\MeV} \cite{miller2015measurement}. This is in tension with evaluations based on extrapolations from the $^{238}$U(p, f) cross section that recommend an increasing cross section to a constant value of roughly \SI{1.5}{\barn} at 1 GeV \cite{duran2017search,carlson2018evaluation}. We have used the LANSCE cross section and assumed a 5\% systematic uncertainty below \SI{200}{\MeV}, a 10\% uncertainty between 200 and \SI{300}{\MeV}, and a constant 20\% uncertainty between 300 and \SI{750}{\MeV}. The uncertainty in the neutron energy spectrum due to the timing uncertainty in the TOF measurement (FWHM $\sim$ \SI{1.2}{\nano\second}) is included in all calculations but is sub-dominant (2.5\%-3.5\%) for the estimates of isotope production rates.

While the nominal beam diameter was set by the 1" collimator, the cross-sectional beam profile has significant tails at larger radii. At the fission chamber approximately 38.8\% of neutrons fall outside a 1" diameter, as calculated with the beam profile provided by LANSCE. % 0.612205000246 in 1" disk, 0.695670261143 in 1" by 1" square, 0.841986246843 in 2" disk. These are calculated using 1inch_Beam_Profile.root on github. ccd: 0.831076034101, fission chamber: 0.999986744216.
Additionally the beam is slightly diverging, with an estimated cone opening angle of 0.233\degree. A Geant4 \cite{agostinelli2003geant4,allison2016recent} simulation that included the measured beam profile and beam divergence, the measured neutron spectrum, and the full geometry and materials of the targets, mounting apparatus, and fission chamber, was used to calculate the neutron fluence through each material, accounting for any attenuation of the neutrons through the targets.  
To reduce computational time, a biasing technique was used to generate neutrons. Instead of following the beam profile, neutrons were generated uniformly in a \SI{16}{\cm}$\times$\SI{16}{\cm} square in front of the fission chamber, covering the entire cross-sectional area of the setup. After running the Geant4 simulation, each event was assigned a weight which is proportional to the intensity of the beam at the simulated neutron location, as obtained from the two-dimensional beam profile supplied by LANSCE. This allows reuse of the same simulation results for different beam profiles and alignment offsets. A total of \num{5.5 E10} neutrons above 10 MeV were simulated for each setup and physics list.
% 960 subruns * 40M * 3 = 1.15E11
%>10 MeV: 474589/1e6
At this level of statistics, the statistical uncertainties in the simulation are sub-dominant to the total neutron fluence uncertainty.

The simulations show that each CCD receives about \SI{83}{\percent} of the whole beam. To assess the uncertainty in the neutron fluence due to misalignment of the beam with the center of the CCDs, the profile of the beam was reconstructed by measuring the dark current rate in the CCDs as a function of position (see Sec.~\ref{sec:ccd_counting}). The beam misalignment is calculated to be about $-2.3$\,mm in the $x$ direction and $+0.5$\,mm in the $y$ direction, which when input into the Geant4 simulation yields a systematic uncertainty in the neutron fluence of less than 1\%. The total neutron fluence ($>$ \SI{10}{\MeV}) through each CCD and its Si-wafer matched pair is listed in Table~\ref{tab:neutron_fluences}; corresponding energy spectra are shown in Fig.~\ref{fig:lanscebeamenergy} (the spectral shape of the fluence through each Si-wafer pair is very similar to that of the corresponding CCD and has been omitted for clarity).

\begin{table}
    \centering
    \begin{tabular}{c c c}
    \hline
    Target & Exposure time & Neutrons through target  \\
        & [hrs] &  ($> 10$ MeV)\\
    \hline
    \vrule width 0pt height 2.2ex
    CCD 1 & 109.4 & \num{2.39 \pm 0.18 E12}\\
    Wafer 1 & 109.4 & \num{2.64 \pm 0.20 E12}\\
    \hline
    \vrule width 0pt height 2.2ex
    CCD 2 & 62.7 & \num{1.42 \pm 0.11 E12}\\
    Wafer 2 & 62.7 & \num{1.56 \pm 0.12 E12}\\
    \hline
    \vrule width 0pt height 2.2ex
    CCD 3 & 22.8 & \num{5.20 \pm 0.39 E11}\\
    Wafer 3 & 22.8 & \num{5.72 \pm 0.43 E11}\\
    \hline
    \end{tabular}
    \caption{Beam exposure details for each CCD and its Si-wafer matched pair.}
    \label{tab:neutron_fluences}
\end{table}

\section{Counting}
\label{sec:counting}
\subsection{Wafers}
\label{ssec:wafer_counting}

\begin{table*}[ht]
\centering
\begin{tabular}{ccccc}
\hline 
 & Wafer 0 & Wafer 1 & Wafer 2 & Wafer 3 \\ 
\hline
\vrule width 0pt height 2.2ex
Si areal density [atoms/cm$^2$] & \multicolumn{4}{c}{\num{4.99 \pm 0.17 e21}~~~~~~~~~~~~~~~~~~~~~} \\
%Beam Neutrons ($> $10 MeV) [$10^{12}$ n] & 0 & \num{2.64 \pm 0.27}& \num{1.56 \pm 0.16} & \num{0.572 \pm 0.058}\\
%Neutron Ratio to Wafer 1 & 0 & 1 & \num{0.5905 \pm 0.0064} & \num{0.2164 \pm 0.0023}\\
Beam to meas.\ time [days] & - & \num{184.107} & \num{187.131} & \num{82.342} \\
Ge counting time [days] & \num{7.000} & \num{1.055}  & \num{3.005} & \num{7.000}  \\
\hline 
\vrule width 0pt height 2.2ex
Measured \ber~activity [mBq] & $<$\num{40}  & \num{161 \pm 24} & \num{75 \pm 12} & \num{149 \pm 12}\\
Decay-corrected \ber~activity [mBq] & - & \num{1830 \pm 270} & \num{870 \pm 140} & \num{437 \pm 34}\\
Beam-avg.\ \ber~cross section [cm$^2$] & - & \num{0.92 \pm 0.16 E-27} & \num{0.74 \pm 0.13 E-27} & \num{1.01 \pm 0.12 E-27}\\
\hline 
\vrule width 0pt height 2.2ex
Measured \sod~activity [mBq] & $<$\num{5.1} & \num{606 \pm 29} & \num{370 \pm 16} &  \num{139.5 \pm 6.3}\\
Decay-corrected \sod~activity [mBq] & - & \num{694 \pm 33} &  \num{424 \pm 19} & \num{148.2 \pm 6.6}\\
Beam-avg.\ \sod~cross section [cm$^2$] & - & \num{6.23 \pm 0.60 E-27} & \num{6.44 \pm 0.61 E-27} & \num{6.15 \pm 0.58 E-27}\\
\hline 
\end{tabular}
\caption{Gamma-counting results for the Si-wafer pairs. Measured activities are corrected for isotope decay that occurred during the beam exposure, as well as between the end of the beam exposure and the time of the gamma counting. Uncertainties are listed at 1$\sigma$ (68.3\%) confidence while upper limits quoted for the unirradiated pair (``Wafer 0'') represent the spectrometer's minimum detectable activity (Currie MDA with a 5\% confidence factor~\cite{currie}) at the corresponding peak energy.}
\label{tab:wafer_counting}
\end{table*}

The gamma-ray activities of the Si-wafer pairs (including the unirradiated pair) were measured with a low-background counter at Pacific Northwest National Laboratory (PNNL). Measurements were performed using a Canberra Broad Energy Ge (BEGe) gamma-ray spectrometer (model BE6530) situated within the shallow underground laboratory (SUL) at PNNL \cite{aalseth2012shallow}. The SUL is designed for low-background measurements, with a calculated depth of \SI{30}{\meter} water equivalent.
%, which results in approximately 100$\times$ fewer fast neutrons and 6$\times$ fewer muons. 
The BEGe spectrometer is optimized for the measurement of fission and activation products, combining the spectral advantages of low-energy and coaxial detectors, with an energy range from \SI{3}{\keV} to \SI{3}{\MeV}.  % and eliminate the requirement for an individual MCA, high-voltage power supply and dedicated coincidence electronics needed for operating a cosmic veto system. 
%The BEGe high voltage was set at +4500 V with a rise time of 8 $\mu$s and flat top of 1 $\mu$s. 
The detector is situated within a lead shield (200\,mm), lined with tin (1\,mm) and copper (1\,mm). It is equipped with a plastic scintillator counter \cite{burnett2017development, burnett2014cosmic, burnett2012development, burnett2013further} to veto cosmic rays, which improves sensitivity by further reducing the cosmic-induced detector background by 25\%. 
The detector was operated with a Canberra Lynx MCA to provide advanced time-stamped list mode functionality.
\begin{figure*}[t!]
\centering
\includegraphics[width=\textwidth]{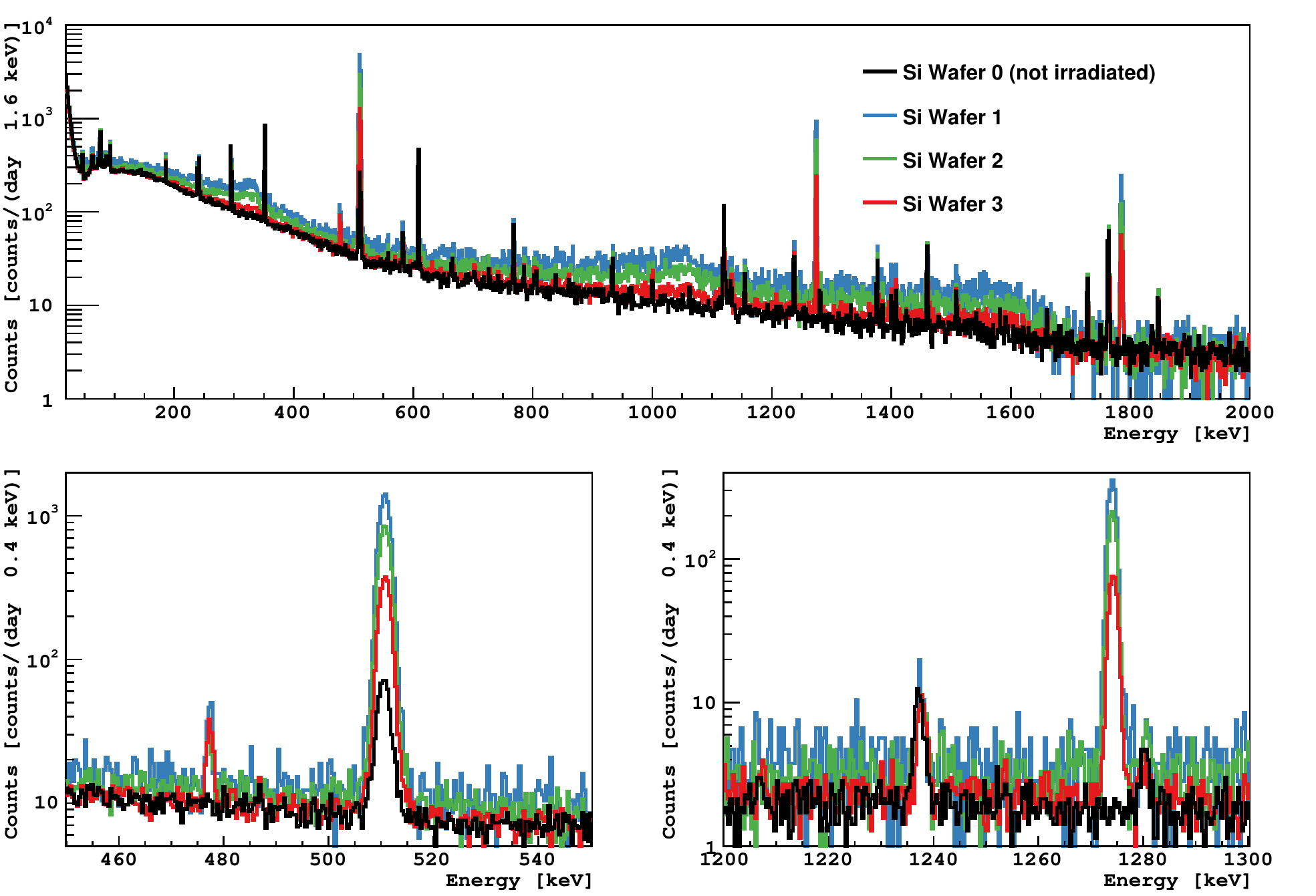}
\caption{Spectral comparison of the gamma-counting results for the Si-wafer pairs.  Inspection of the full energy range (top panel) reveals two peaks in the irradiated samples (1, 2, and 3) at \SI{478}{\keV} (bottom left) and \SI{1275}{\keV} (bottom right) that are not present in the unirradiated sample (0), corresponding to \ber\ and \sod\ activated by the LANSCE neutron beam, respectively.}
\label{fig:ge_counting}
\end{figure*}

Each wafer pair was measured independently, with wafer pair 3 and the unexposed wafer pair 0 counted for longer periods because their expected activities were the lowest. Table~\ref{tab:wafer_counting} shows the gamma-counting details, and Fig.~\ref{fig:ge_counting} shows the measured gamma-ray spectra. Spectral analysis was performed using the Canberra Genie 2000 Gamma Acquisition \& Analysis software (version 3.4) and all nuclear data were taken from the Evaluated Nuclear Data File (ENDF) database \cite{chadwick2011endf} hosted at the National Nuclear Data Center by Brookhaven National Laboratory. Compared to the unirradiated wafer-pair spectrum, the only new peaks identified in the spectra of the irradiated wafer pairs are at 478 and \SI{1275}{\keV}, corresponding to \ber~(10.44\% intensity per decay) and \sod~(99.94\% intensity per decay), respectively (cf.\,Fig.\,\ref{fig:ge_counting}). Note that each of the irradiated wafer pairs also has a significant excess at \SI{511}{\keV}, corresponding to positron-annihilation photons from \sod\ decays, and an associated sum peak at \SI{1786}{\keV} ($= 511 +$ \SI{1275}{\keV}). 

The \ber\ and \sod\ activities in each wafer pair were calculated using the 478 and \SI{1275}{\keV} peaks, respectively. The measured values listed in Table~\ref{tab:wafer_counting} include the detector efficiency and true-coincidence summing corrections for the sample geometry and gamma-ray energies considered (calculated using the Canberra In Situ Object Counting Systems, or ISOCS, calibration software \cite{venkataraman1999validation}). The activity uncertainties listed in Table~\ref{tab:wafer_counting} include both the statistical and systematic contributions, with the latter dominated by uncertainty in the efficiency calibration ($\sim$\SI{4}{\percent}). Each measured activity is then corrected for isotope decay that occurred during the beam exposure, as well as between the end of the beam exposure and the time of the gamma counting. 

To compare among the results of the different wafer pairs, we divide each decay-corrected activity by the total number of incident neutrons and the number of target Si atoms to obtain a beam-averaged cross section (also listed in Table~\ref{tab:wafer_counting}). The values are in good agreement for both \ber\ and \sod\ (even if the common systematic uncertainty associated with the neutron beam fluence is ignored), which serves as a cross-check of the neutron-beam exposure calculations. The lack of any other identified peaks confirms that there are no other significant long-lived gamma-emitting isotopes produced by high-energy neutron interactions in silicon. Specifically, the lack of an identifiable peak at \SI{1808.7}{\keV} allows us to place an upper limit on the produced activity of $^{26}$Al at the minimum detectable activity level of \SI{12}{\milli\becquerel} (Currie MDA with a 5\% confidence factor~\cite{currie}), i.e.\ at least 58$\times$ lower than the \sod\ activity in wafer pair 1.

\subsection{CCDs}
\label{sec:ccd_counting}
Images from CCD\,3 were acquired at The University of Chicago in a custom vacuum chamber. Prior to counting, the CCD was removed from the aluminum transport box and placed in a copper box inside the vacuum chamber. Images taken were 4200 columns by 2100 rows in size, with 52 rows and 104 columns constituting the ``overscan'' (i.e., empty pixel reads past the end of the CCD pixel array). These overscan pixels contain no charge and thus provide a direct measurement of the pixel readout noise. A total of 8030 post-irradiation images with \SI{417}{\sec} of exposure were acquired, for a total counting time of 38.76 days. Data were taken in long continuous runs of many images, with interruptions in data taking for testing of the CCD demarcating separate data runs.

Background data were taken prior to shipment to the LANSCE facility for neutron irradiation. These background data consist of the combined spectrum from all radioactive backgrounds in the laboratory environment, including the vacuum chamber, the intrinsic contamination in the CCD, and cosmic rays. A total of 1236 images were acquired using the same readout settings as post-irradiation images, but with a longer exposure of \SI{913}{\sec}, for a total counting time of 13.06 days.

\begin{figure}
    \centering
    \includegraphics[width=\columnwidth]{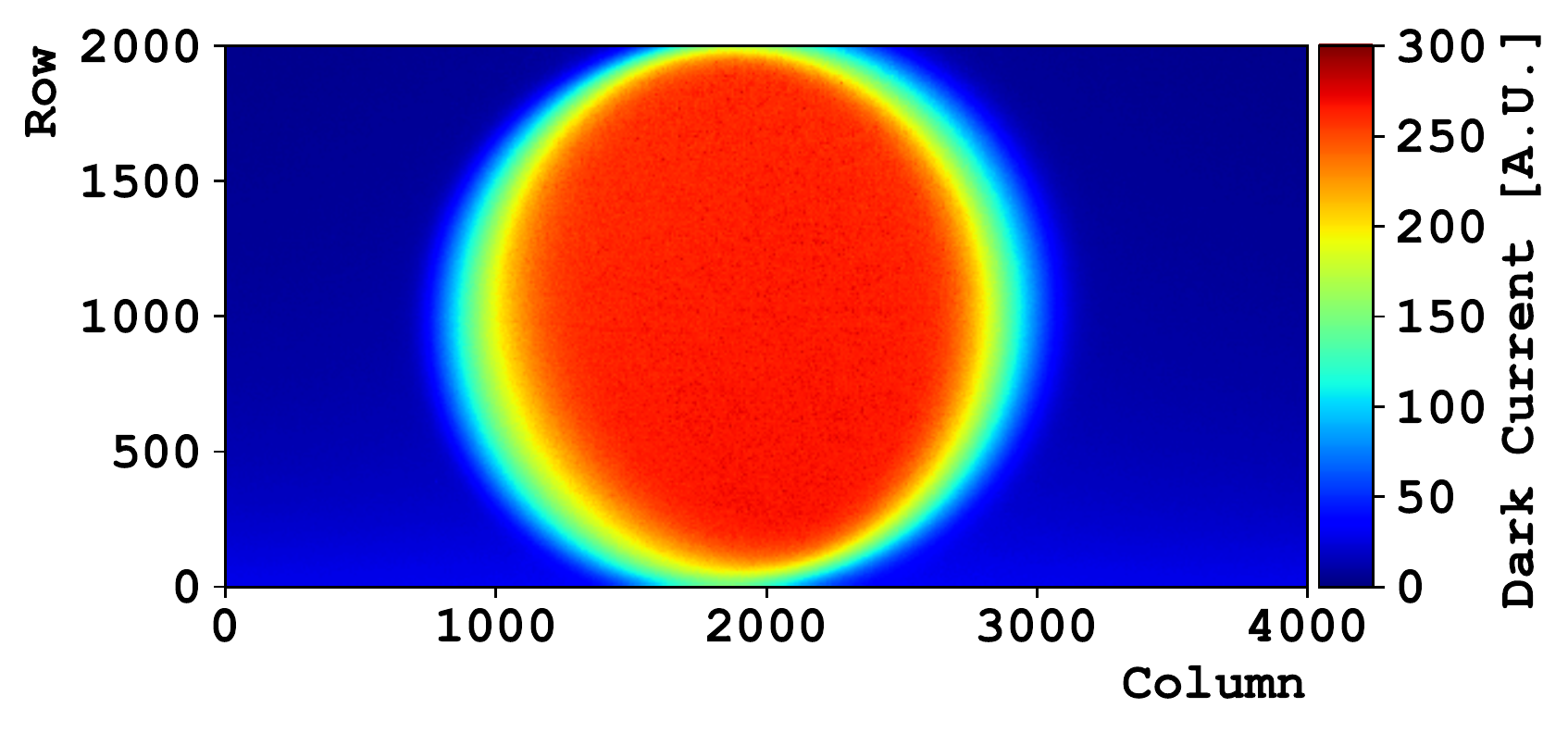}
    \caption{Post-irradiation dark-current profile for CCD\,3, obtained from the median pixel values across multiple images. The elevated number of dark counts in the center of the CCD shows the effect of the neutron damage on the CCD.}
    \label{fig:darkcurrentprofile}
\end{figure}

CCD images were processed with the standard DAMIC analysis software~\cite{PhysRevD.94.082006}, which subtracts the image pedestal, generates a ``mask'' to exclude repeating charge patterns in the images caused by defects, and groups pixels into clusters that correspond to individual ionization events. The high dark current caused by damage to the CCD from the irradiation (see Fig.~\ref{fig:darkcurrentprofile}) necessitated a modification to this masking procedure because the average CCD pixel values were no longer uniform across the entire CCD, as they were before irradiation. The images were therefore split into 20-column segments which were treated separately for the pedestal subtraction and masking steps.

Simulations of \trit{}, \sod{}, and \ber{} decays in the bulk silicon of the CCD were performed using a custom Geant4 simulation, using the Penelope Geant4 physics list, with a simplified geometry that included only the CCD 
%TODO: maybe add comment about CCD thickness/dead layers?
and the surrounding copper box. Radioactive-decay events were simulated according to the beam profile, assumed to be proportional to the dark current profile (shown in Fig. ~\ref{fig:darkcurrentprofile}). The CCD response was simulated for every ionization event, including the stochastic processes of charge generation and transport that were validated in Ref.~\cite{PhysRevD.96.042002}.

%\ber{} was also simulated to determine if it would be a significant background in the final spectrum and was determined to have a negligible contribution to the final spectrum even if some amount of it was still present. See figure \ref{fig:comparisonbe7na22}.
To include the effects of noise and dark current on the clustering algorithm, simulated ``blank'' images were created with the same noise and dark-current profile as the post-irradiation data. The simulated ionization events were pixelated and added onto the blank images, which were then processed with the standard DAMIC reconstruction code to identify clusters.
The increase in the vertical (row-to-row) charge transfer inefficiency (CTI) observed in the post-irradiation data was simulated with a Poissonian kernel, which assumes a constant mean probability, $\lambda$, of charge loss for each pixel transfer along a column~\cite{janesick}. We assume a dependence of $\lambda$ as a function of column number that is proportional to the dark current profile. The total effect of CTI on a particular cluster depends on the number of vertical charge transfers $n$. The continuous CCD readout scheme, chosen to optimize the noise while minimizing overlap of charge clusters, results in a loss of information about the true number of vertical charge transfers for each cluster. For every simulated cluster we therefore pick a random $n$ uniformly from 1 to 2000 to simulate events distributed from the bottom row to the top row of the CCD and apply the Poissonian kernel. We determined the maximum value of $\lambda$ near the center of the CCD to be $9\times10^{-4}$ by matching the distribution of the vertical spread of clusters in the simulation to the data.\footnote{The data from CCD\,1 and CCD\,2, which experienced significantly higher neutron irradiation than CCD\,3, were discarded from the analysis because the vertical CTI could not be well described with a Poissonian kernel. We suspect that the CTI in these CCDs is dominated by the effect of charge traps introduced by the neutron irradiation. During the readout procedure these traps are filled with charge from ionization clusters. The charge is then released on the time scale of milliseconds, corresponding to $\sim$25 vertical transfers. This effect is difficult to model and results in considerable loss of charge from clusters in these two CCDs.}

The identified clusters in the background data acquired prior to irradiation at LANSCE were also introduced on simulated blank images to include the effect of dark current, defects, and CTI on the background spectrum in the activated region of the CCD.

The post-irradiation energy spectrum was fit using a model that includes components for the CCD background, \sod{} decays, and \trit{} decays. \ber{} was excluded from the fit because the decay does not produce a significant contribution to the total energy spectrum, even if the activity were many times the value we expect based on the wafer measurement. 

\begin{figure}
    \centering
    \includegraphics[width=\columnwidth]{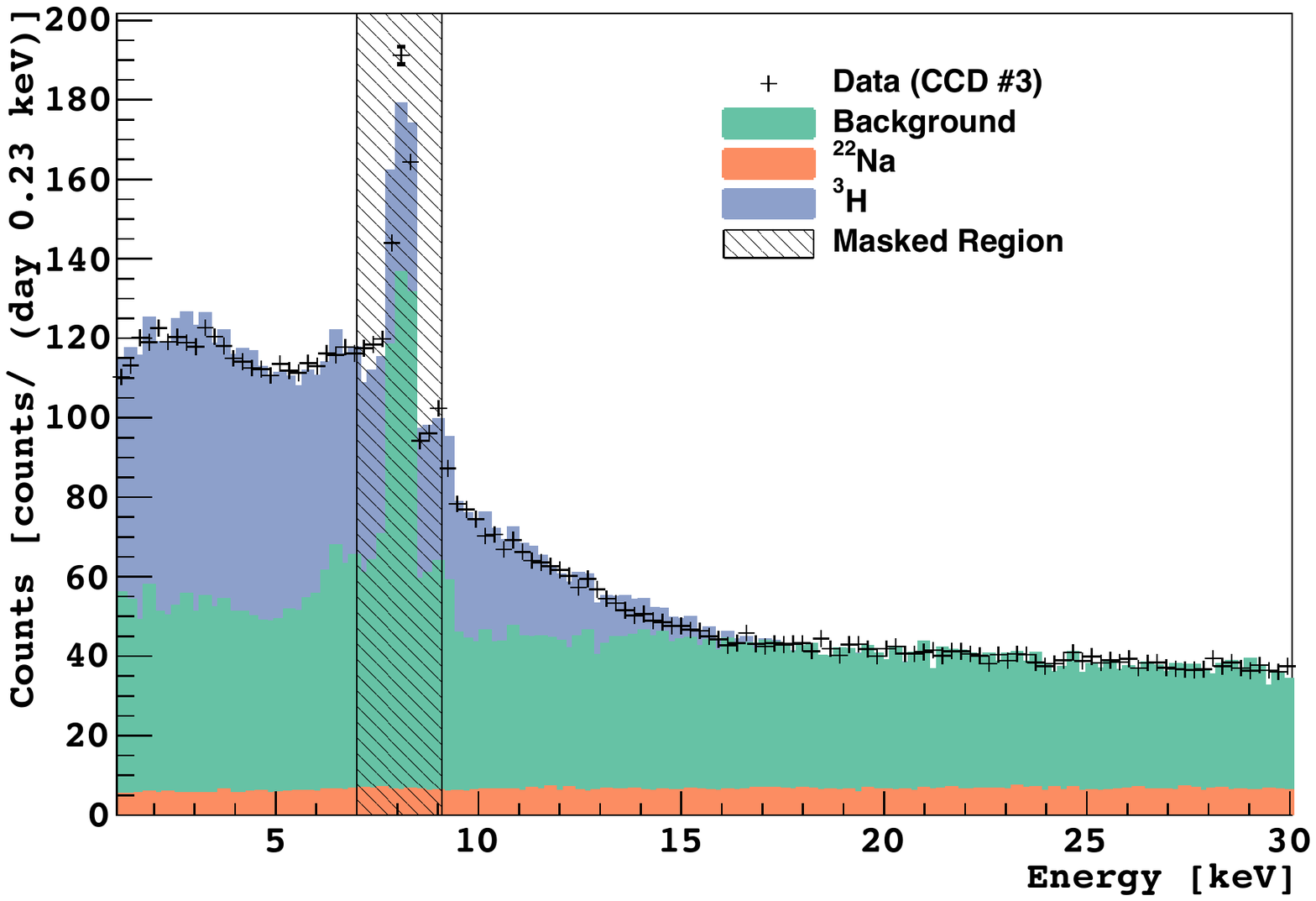}
    \caption{Data spectrum and best-fit model with the spectral components stacked in different colors. The spectrum was fit from 2 to \SI{25}{\keV} with the shaded region around the \SI{8}{\keV} copper K-shell fluorescence line excluded from the fit. The rise in the spectrum below \SI{18}{\keV} from \trit{} decay is clearly visible above the nearly flat background and \sod{} spectrum.}
    \label{fig:finalfitresults}
\end{figure}

We constructed a binned Poissonian log-likelihood as the test statistic for the fit, which was minimized using Minuit \cite{James:1994vla} to find the best-fit parameters.
Due to the relatively low statistics in the background template compared to post-irradiation data, statistical errors were corrected using a modified Barlow-Beeston method \cite{BARLOW1993219}, allowing each bin of the model to fluctuate by a Gaussian-constrained term with a standard deviation proportional to the bin statistical uncertainty.
%This produced no change in the fit results but allows a correct estimation of the statistical uncertainties.
The data spectrum was fit from 2 to \SI{25}{\kilo\eV} to contain most of the \trit{} spectrum, while excluding clusters from noise at low energies.
A \SI{2}{\kilo\eV}-wide energy region around the copper K-shell fluorescence line at \SI{8}{\kilo\eV} was masked from the fit because it is not well-modeled in the simulation.
This peak-like feature is more sensitive to the details of the energy response than the smooth \trit{} spectrum. We have verified that including this K-shell line in the fit has a negligible effect on the fitted \trit\ activity.
The background rate for the fit was fixed to the pre-irradiation value, while keeping the amplitude of the \sod{} spectrum free.
This choice has a negligible impact on the \trit{} result because the background and \sod{} spectra are highly degenerate within the fit energy range, with a correlation coefficient of 0.993.
%issues with cluster reconstruction due to damage to the CCD prevent breaking the degeneracy with higher energy events,
%This choice has no significant effect on the \trit{} result because its value is driven primarily by the shape of the spectrum independently of the flat background.
Figure~\ref{fig:finalfitresults} shows the measured energy spectrum and the best-fit result ($\chi^2$/NDF=104/87).

After the fit was performed, the activities were calculated by dividing the fitted counts by the cumulative data exposure. This number was corrected for the isotope-specific event detection efficiency obtained from the simulation for the energy region of interest.  
% (necessary because our detector may reconstruct more or less than one cluster per decay for \sod{} and \trit{}, depending on the behavior of the emitted $\beta$ particle).
%after correcting for inefficiencies using the Geant4 simulations.
Systematic errors were estimated from a series of fits under different configurations, including varying the energy range of the fit, varying the energy response and charge transfer parameters within their uncertainties, and floating versus constraining the amplitudes of the background and/or \sod{} components in the fit. 
The best estimate for the tritium activity in CCD\,3 (after correcting for radioactive decay) is $45.7 \pm 0.5 $ (stat) $\pm 1.5 $ (syst) \si{\milli\becquerel}.
%$126 \pm 5 $ (stat) $ \pm 25 $ (syst) mBq for \sod{}, and $45.7 \pm 0.5 $ (stat) $\pm 1.5 $ (syst) mBq for \trit{}.

%during the data taking period for CCD\,3 were $118 \pm 5$ (stat) $ \pm 5 $ (syst) mBq for \sod{}, and $45.0 \pm 0.5  $ (stat) $ \pm 1.5 $ (syst) mBq for \trit{}. Correcting for radioactive decay between irradiation and measurement gives values of $126 \pm 5 $ (stat) $ \pm 5 $ (syst) mBq for \sod{}, and $45.7 \pm 0.5 $ (stat) $\pm 1.5 $ (syst) mBq for \trit{} at the time of irradiation.

The precision of the \sod\ measurement in the CCDs is limited because the relatively flat \sod{} spectrum is degenerate with the shape of the background spectrum. Unfortunately, there are no features in the CCD spectrum at low energies that can further constrain the \sod{} activity. Further, the damage to the CCD renders the spectrum at higher energies unreliable because events with energies $>$\SI{50}{\kilo\eV} create large extended tracks where the effects of CTI, dark current, and pileup with defects become considerable, preventing reliable energy reconstruction. 
Notably, characteristic full-absorption $\gamma$ lines are not present in the CCD spectrum because $\gamma$ rays do not deposit their full energy in the relatively thin CCDs. As a cross-check of the post-irradiation background rate, we separately fit the first and last 400 columns of the CCD (a region mostly free of neutron exposure) and found values consistent with the pre-irradiation background to within $\sim$\SI{7}{\percent}. Constraining the background to within this range has a negligible effect on the fitted tritium activity but leads to significant variation in the estimated \sod\ activity, which dominates the overall systematic uncertainty. The best estimate for the \sod~activity in CCD\,3 is $126 \pm 5 $ (stat) $ \pm 26 $ (syst) \si{\milli\becquerel}. This is consistent with the more precise measurement of the \sod~activity in the silicon wafers, which corresponds to a CCD\,3 activity of \SI{88.5 \pm 5.3}{\milli\becquerel}.

\section{Predicted Beam Production Rate}
\label{sec:production_rates}
If the neutron beam had an energy spectrum identical to that of cosmic-ray neutrons,  we could simply estimate the cosmogenic production rate by scaling the measured activity by the ratio of the cosmic-ray neutrons to that of the neutron beam. However the beam spectrum falls off faster at higher energies than that of cosmic rays (see Fig.~\ref{fig:lanscebeamenergy}). Thus we must rely on a model for the production cross sections to extrapolate from the beam measurement to the cosmogenic production rate. 

We can evaluate the accuracy of the different cross-section models by comparing the predicted \trit, \ber, and \sod~activity produced by the LANSCE neutron beam irradiation to the decay-corrected measured activities. We note that measurements of the unirradiated targets confirm that any non-beam related isotope concentrations (e.g. due to cosmogenic activation) are negligible compared to the beam-induced activity.
%The number of atoms, $N(t)$, of a given isotope at a given time $t$ is governed by the equation
%\begin{linenomath*}
%\begin{align}
%\frac{dN}{dt} = + A(t) -\frac{N(t)}{\tau}
%\end{align}
%\end{linenomath*}
%where $\tau$ is the mean life [\si{\second}] of the isotope decay and $A(t)$ is the isotope production rate [\si{\atoms\per \second}]. Ignoring any isotope production before ($t<0$) or after ($t>t_b$) the beam exposure (found to be negligible from measurements of the unirradiated targets), the measured decay rate, $D(t)$, at any time after the end of the beam is given by
%\begin{linenomath*}
%\begin{align}
%D(t) = \frac{N(t)}{\tau} = \frac{1}{\tau}\int\limits_0^{t_b} %A(t')e^{\frac{-(t-t')}{\tau}}dt'
%\end{align}
%\end{linenomath*}
For a given model of the isotope production cross section $\sigma(E)$ [cm$^2$], the predicted isotope activity, $P$ [Bq], produced by the beam (correcting for decays) is given by
\begin{linenomath*}
\begin{align}
\label{eq:beam_act}
P = \frac{n_a}{\tau} \int S(E) \cdot \sigma(E)~dE
\end{align}
\end{linenomath*}
where $n_a$ is the areal number density of the target silicon atoms [\si{\atoms\per \cm\squared}], $\tau$ is the mean life [\si{\second}] of the isotope decay, and $S(E)$ is the energy spectrum of neutrons [\si{\neutrons \per \MeV}]. The second column of Table~\ref{tab:trit_pred} shows the predicted activity in CCD 3, $P_\text{CCD3}$, for the different \trit~cross-section models considered. The corresponding numbers for \ber~and \sod~in Wafer 3 ($P_\text{W3})$ are shown in Tables~\ref{tab:ber_pred} and  \ref{tab:sod_pred} respectively. The uncertainties listed include the energy-dependent uncertainties in the LANSCE neutron beam spectrum and the uncertainty in the target thickness.

\begin{table*}[t!]
   \centering
   \begin{tabular}{cccccc} 
   \hline
   Model & Predicted LANSCE & Ejected & Implanted & Predicted LANSCE & Measured/Predicted\\
   & \trit~produced act. & activity & activity & \trit~residual act. & \trit~residual activity\\
   & $P_\text{CCD3}$ [\si{\milli\becquerel}] & $E_\text{CCD3}$ [\si{\milli\becquerel}] & $I_\text{CCD3}$ [\si{\milli\becquerel}] & $R_\text{CCD3}$ [\si{\milli\becquerel}] & \\
   \hline
   K\&K (ACTIVIA) & \num{40.8 \pm 4.5} & & &\num{41.5 \pm 5.6} & \num{1.10 \pm 0.15}\\
   TALYS & \num{116 \pm 16} & \num{46.70 \pm 0.12} & \num{53.8 \pm 2.1} & \num{123 \pm 17} & \num{0.370 \pm 0.053} \\
   %TALYS & \num{119 \pm 16} & \num{0.403 \pm 0.001} & \todo{\num{26.0 \pm 37.1}} & \num{96.9 \pm 37.3} \\
   INCL++(ABLA07) & \num{41.8 \pm 4.8} & & & \num{42.5 \pm 5.9} & \num{1.07 \pm 0.15}\\
   GEANT4 BERTINI & \num{13.0 \pm 1.5} & \num{3.354 \pm 0.072} & \num{3.699 \pm 0.045} & \num{13.3 \pm 1.6} & \num{3.43 \pm 0.42}\\
   GEANT4 BIC & \num{17.8 \pm 1.8} & \num{4.995 \pm 0.084} & \num{6.421 \pm 0.059} & \num{19.2 \pm 2.0} & \num{2.38 \pm 0.26}\\
   GEANT4 INCLXX & \num{42.3 \pm 5.1} & \num{20.65 \pm 0.11} & \num{16.94 \pm 0.10} & \num{38.5 \pm 4.6} & \num{1.19 \pm 0.15}\\
   \hline
   \end{tabular}
   \caption{Predicted \trit~activity in CCD 3 based on different cross-section models. The second column lists the total predicted activity produced in the CCD. The third and fourth columns list the activity ejected and implanted respectively with listed uncertainties only due to simulation statistics. The fifth column shows the final predicted residual activity calculated from the second, third, and fourth columns, including systematic uncertainties due to the geometry. For models without ejection and implantation information we use the average of the other models---see text for details. The final column shows the ratio of the experimentally measured activity to the predicted residual activity.}
   \label{tab:trit_pred}
\end{table*}

\begin{table*}[t!]
   \centering
   \begin{tabular}{cccccc} 
   \hline
   Model & Predicted LANSCE & Ejected & Implanted & Predicted LANSCE & Measured/Predicted\\
   & \ber~produced act. & activity & activity & \ber~residual act. & \ber~residual act.\\
   & $P_\text{W3}$ [\si{\milli\becquerel}] & $E_\text{W3}$ [\si{\milli\becquerel}] & $I_\text{W3}$ [\si{\milli\becquerel}] &  $R_\text{W3}$ [\si{\milli\becquerel}] & \\
   \hline
   S\&T (ACTIVIA) & \num{408 \pm 46} & & & \num{405 \pm 49} & \num{1.08 \pm 0.16}\\
   TALYS & \num{294 \pm 41} & & & \num{292 \pm 42} & \num{1.50 \pm 0.25}\\
   INCL++(ABLA07) & \num{141 \pm 21} & & & \num{140 \pm 22} & \num{3.12 \pm 0.55}\\
   $^{\text{nat}}$Si(p,x)$^7$Be Spline Fit & \num{518 \pm 68} & & & \num{514 \pm 72} & \num{0.85 \pm 0.14}\\
   %GEANT4 BERTINI & \num{0.99 \pm 0.19} & \num{0.00 \pm 0.20} & \num{0.64 \pm 0.14} & \num{1.63 \pm 0.43} & \num{268 \pm 73} \\
   %GEANT4 BIC & \num{1.27 \pm 0.24} & \num{0.00 \pm 0.20} & \num{0.61 \pm 0.16} & \num{1.98 \pm 0.50} & \num{221 \pm 58}\\
   GEANT4 BERTINI & \num{0.99 \pm 0.20} & $<0.33$ & \num{0.64 \pm 0.14} & \num{1.63 \pm 0.43} & \num{268 \pm 74} \\
   GEANT4 BIC & \num{1.27 \pm 0.24} & $<0.33$ & \num{0.61 \pm 0.16} & \num{1.98 \pm 0.50} & \num{221 \pm 59}\\
   GEANT4 INCLXX & \num{21.6 \pm 3.0} & \num{3.59 \pm 0.85} & \num{3.42 \pm 0.38} & \num{21.4 \pm 3.1} & \num{20.4 \pm 3.4}\\
   \hline
   \end{tabular}
   \caption{Predicted \ber~activity in Wafer 3 based on different cross-section models. See Table~\ref{tab:trit_pred} caption for a description of the columns. Upper limits are 90\% C.L.} % UL = CV + 1.64 error
   \label{tab:ber_pred}
\end{table*}

\begin{table*}[t!]
   \centering
   \begin{tabular}{cccccc} 
   \hline
   Model & Predicted LANSCE & Ejected & Implanted & Predicted LANSCE & Measured/Predicted\\
   & \sod~produced act. & activity & activity & \sod~residual act. & \sod~residual act.\\
   & $P_\text{W3}$
   [\si{\milli\becquerel}] & $E_\text{W3}$ [\si{\milli\becquerel}] & $I_\text{W3}$ [\si{\milli\becquerel}] & $R_\text{W3}$ [\si{\milli\becquerel}] & \\
   \hline
   S\&T (ACTIVIA) & \num{295 \pm 29} &  &  & \num{295 \pm 29} & \num{0.502 \pm 0.054}\\
   TALYS & \num{209 \pm 18}&  &  & \num{208 \pm 18} & \num{0.711 \pm 0.070}\\
   INCL++(ABLA07) & \num{207 \pm 21} &  &  & \num{206 \pm 21} & \num{0.718 \pm 0.081}\\
   Michel-TALYS & \num{151 \pm 14} &  &  & \num{151 \pm 14} & \num{0.98 \pm 0.10}\\
   %GEANT4 BERTINI & \num{97 \pm 14} & \num{0.0011 \pm 0.0084} & \num{7.04 \pm 4.78 E-4} & \num{97 \pm 14} & \num{1.53 \pm 0.23}\\ % rev1
   %GEANT4 BIC & \num{393 \pm 55} & \num{0.0008 \pm 0.0042} & \num{2.64 \pm 1.87 E-4} & \num{393 \pm 55} & \num{0.377 \pm 0.056}\\ % rev1
   %GEANT4 INCLXX & \num{398 \pm 56} & \num{0.0007 \pm 0.0041} & \num{1.32 \pm 1.32 E-4} & \num{398 \pm 56} & \num{0.372 \pm 0.055}\\ % rev1
   %GEANT4 BERTINI & \num{97 \pm 10} & \num{0.09 \pm 0.48} & \num{0.003 \pm 0.003} & \num{96 \pm 10} & \num{1.54 \pm 0.18}\\ % rev2
   %GEANT4 BIC & \num{393 \pm 39} & \num{0.42 \pm 0.96} &\num{0.008 \pm 0.005} & \num{392 \pm 39} & \num{0.378 \pm 0.041}\\ % rev2
   %GEANT4 INCLXX & \num{398 \pm 39} & \num{0.38 \pm 0.96} & \num{0.002 \pm 0.016} & \num{398 \pm 39} & \num{0.373 \pm 0.040}\\ % rev2
   GEANT4 BERTINI & \num{97 \pm 11} & $< 0.88$ & $<0.008$ & \num{96 \pm 11} & \num{1.54 \pm 0.18}\\
   GEANT4 BIC & \num{393 \pm 40} & $<2.0$ & $<0.02$ & \num{392 \pm 40} & \num{0.378 \pm 0.042}\\
   GEANT4 INCLXX & \num{398 \pm 40} & $<2.0$ & $<0.03$ & \num{398 \pm 40} & \num{0.373 \pm 0.041}\\
   \hline
   \end{tabular}
   \caption{Predicted \sod~activity in Wafer 3 based on different cross-section models. See Table~\ref{tab:trit_pred} caption for a description of the details. Upper limits are 90\% C.L.} % UL = CV + 1.64 error
   \label{tab:sod_pred}
\end{table*}

\begin{figure}
\centering
\includegraphics[width=0.75\columnwidth]{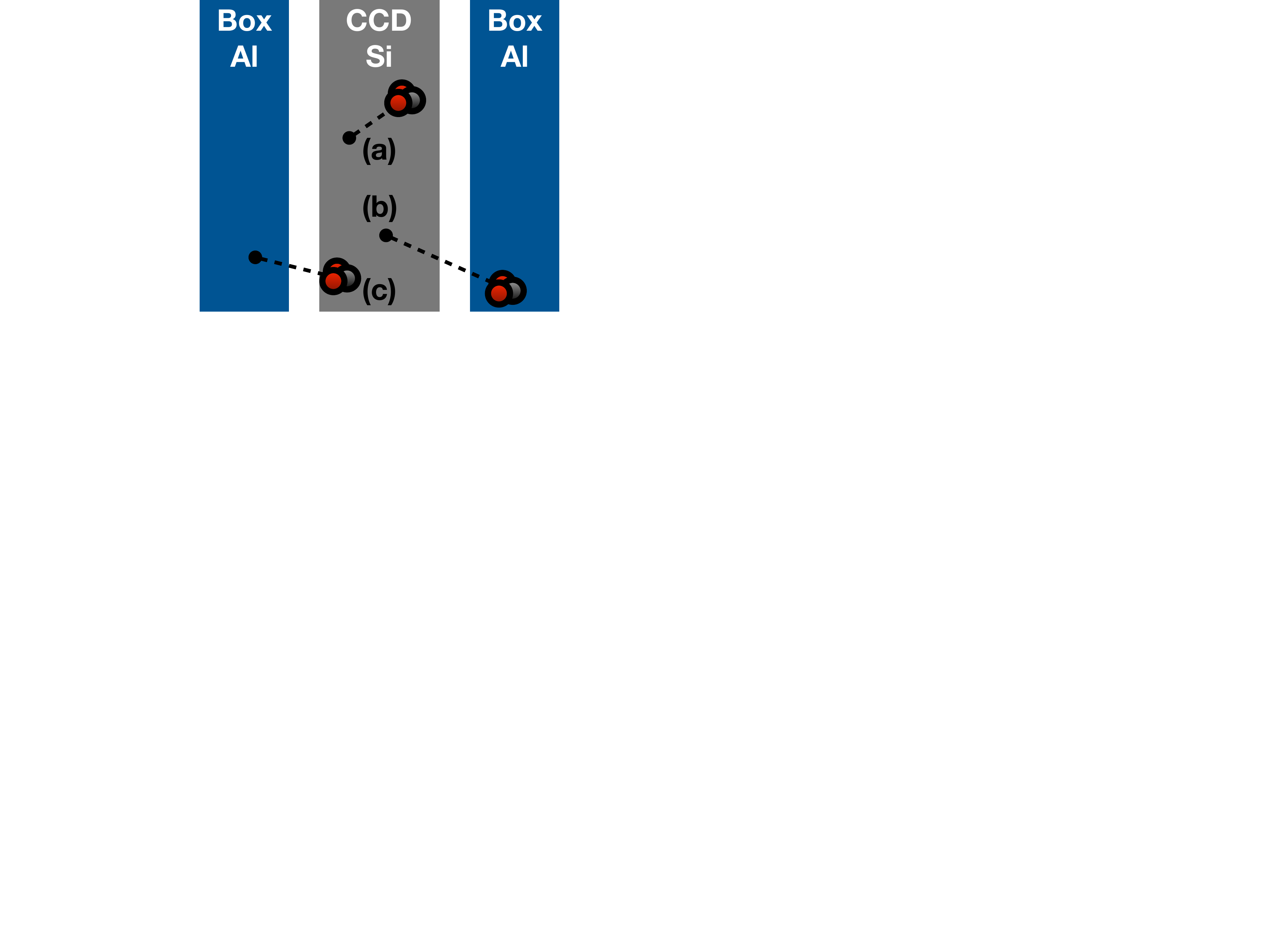}
\caption{Schematic diagram showing triton ejection and implantation. The filled circles indicate example triton production locations while the triton nuclei show the final implantation locations. Production rate estimates include trajectories (a) and (b), while counting the tritium decay activity in the CCD measures (a) and (c).}
\label{fig:trit_ejec_schematic}
\end{figure}

\begin{figure*}
\centering
\includegraphics[width=\textwidth]{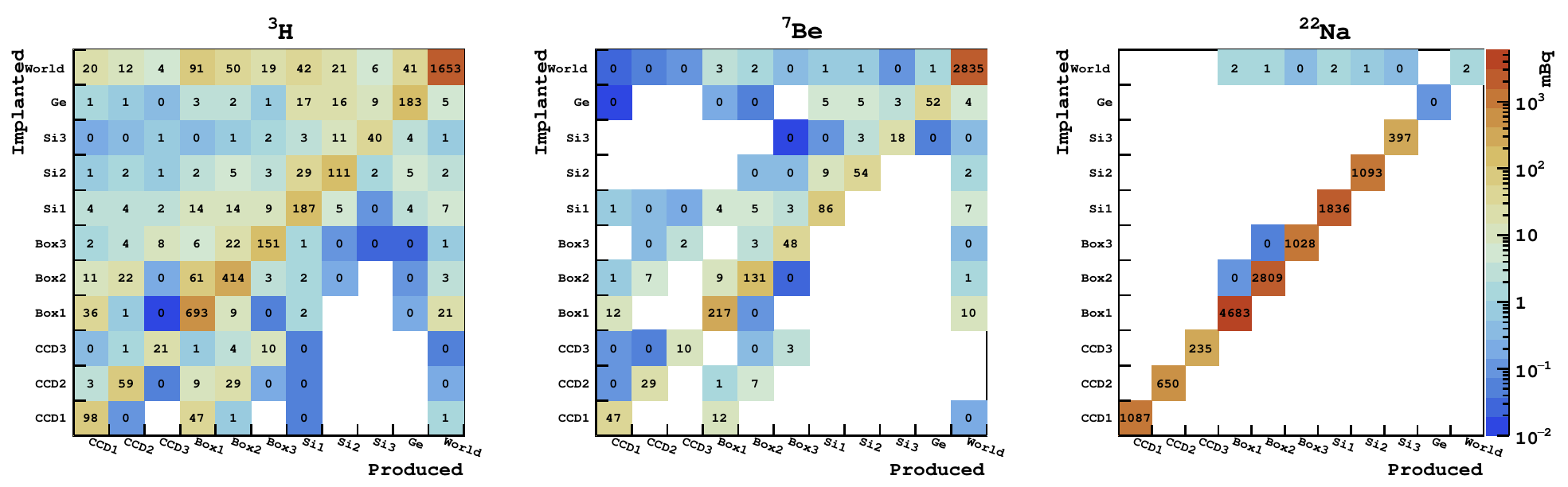}
\caption{Shown are the activities [mBq] of $^3$H (left), $^7$Be (middle), and $^{22}$Na (right) produced and implanted in various volumes  (i.e., $T_{ij}\cdot P_j$) as predicted by the GEANT4 INCLXX model. CCD\,1, CCD\,2, CCD\,3 are the CCDs, with CCD\,1 being closest to the fission chamber. Box\,1, Box\,2, and Box\,3 are the aluminum boxes that contain CCD\,1, CCD\,2, and CCD\,3, respectively. Si\,1, Si\,2, Si\,3, and Ge are the silicon and germanium wafers downstream of the CCDs. World represents the air in the irradiation room.}
\label{fig:transmat}
\end{figure*}

\subsection{Ejection and Implantation}
%There is a complication for light nuclei that was not fully appreciated in the design phase of the experiment.
Light nuclei, such as tritons, can be produced with significant fractions of the neutron kinetic energy. Due to their small mass, these nuclei have relatively long ranges and can therefore be ejected from their volume of creation and implanted into another volume. The situation is shown schematically in Fig.~\ref{fig:trit_ejec_schematic}. While we would like to estimate the total production rate in the silicon targets, what is actually measured is a combination of the nuclei produced in the target that are not ejected and nuclei produced in surrounding material that are implanted in the silicon target. The measured activity therefore depends not only on the thickness of the target but also on the nature and geometry of the surrounding materials.

The residual activity, $R_i$, eventually measured in volume $i$, can be written as
\begin{align}
\label{eq:transfer}
R_i = \sum_j T_{ij} \cdot P_j 
\end{align}
where $P_j$ is the total activity produced in volume $j$ (see Eq.~\ref{eq:beam_act}) and $T_{ij}$ is the transfer probability---the probability of a triton produced in $j$ to be eventually implanted in $i$. Because the ejection and implantation of light nuclei is also an issue for dark matter detectors during fabrication and transportation, we have also explicitly factored the transfer probability into ejected activity ($E_i$) and activity implanted from other materials ($I_i$) to give the reader an idea of the relative magnitudes of the two competing effects: 

\begin{align}
\label{eq:ejection}
E_i &= (1 - T_{ii})\cdot P_i\\
\label{eq:implantation}
I_i &=  \sum_{j \neq i} T_{ij} \cdot P_j\\
R_i &= P_i - E_i + I_i 
\end{align}

%(Note that this also takes into account the staggered beam time.)
For nuclear models that are built-in as physics lists within Geant4, explicit calculations of transfer probabilities are not necessary, because the nuclei produced throughout the setup are propagated by Geant4 as part of the simulation. For the TALYS model, which does calculate the kinematic distributions for light nuclei such as tritons but is not included in Geant4, we had to include the propagation of the nuclei separately. Since the passage of nuclei through matter in the relevant energy range is dominated by electromagnetic interactions, which are independent of nuclear production models and can be reliably calculated by Geant4, we used TALYS to evaluate the initial kinetic energy and angular distributions of triton nuclei produced by the LANSCE neutron beam and then ran the Geant4 simulation starting with nuclei whose momenta are drawn from the TALYS-produced distributions. For the remaining models which do not predict kinematic distributions of the resulting nuclei, we simply used the average and standard deviation of the transfer probabilities from the models that do provide this information. As an example, the transfer matrix (expressed in terms of activity $T'_{ij} = T_{ij}\cdot P_j$) from the Geant4 INCLXX model for all three isotopes of interest is shown in Fig.~\ref{fig:transmat}. The uncertainties are calculated by propagating the statistical errors from the simulations through Eqs.~(\ref{eq:transfer}), (\ref{eq:ejection}), and (\ref{eq:implantation}). Additionally we have evaluated a 1\% systematic uncertainty on ejection and implantation of \trit{} and \ber~due to the uncertainty in the target thicknesses.

\subsubsection{Tritium}
The model predictions for the ejected and implantated activity of tritons in CCD 3 are shown in the third and fourth columns of Table~\ref{tab:trit_pred}. One can see that depending on the model, 25\%--50\% of the tritons produced in the CCDs are ejected and there is significant implantation of tritons from the protective aluminum boxes surrounding the CCDs. 
Due to the similarity of the aluminum and silicon nucleus and the fact that the reaction Q-value for triton production only differs by \SI{5.3}{MeV}, at high energies the production of tritons in aluminum is very similar to that of silicon. In Ref.~\cite{benck2002secondary}, the total triton production cross section as well as the single and double differential cross sections for neutron-induced triton ejection were found to be the same for silicon and aluminum, within the uncertainty of the measurements. This led the authors to suggest that results for aluminum, which are more complete and precise, can also be used for silicon. We show all existing measurements for neutron- and proton-induced triton production in aluminum \cite{benck2002fast, otuka2014towards, zerkin2018experimental} in Fig.~\ref{fig:al_3h_cross_sections} along with model predictions. Comparison to Fig.~\ref{fig:si_3h_cross_sections} shows that all models considered have very similar predictions for aluminum and silicon.

This similarity in triton production, as well as the similar stopping powers of aluminum and silicon, leads to a close compensation of the triton ejected from the silicon CCD with the triton implanted into the CCD from the aluminum box. If the material of the box and CCD were identical and there was sufficient material surrounding the CCD, the compensation would be exact, with no correction to the production required (ignoring attenuation of the neutron flux). In our case, the ratio of production to residual tritons is predicted to be \num{0.985 \pm 0.078}, based on the mean and RMS over all models with kinematic information, and we apply this ratio to the rest of the cross-section models.

\subsubsection{\ber}
Due to the heavier nucleus, the fraction of ejected \ber~nuclei is expected to be smaller than for tritons. As listed in Table~\ref{tab:ber_pred}, the Geant4 INCLXX model predicts that $\sim17\%$ of \ber~produced in the silicon wafers is ejected. For the BIC and BERTINI models, the predicted production rates in silicon are roughly 400 times smaller than our measurement and within the statistics of our simulations we could only place upper limits on the fraction ejected from the wafers at roughly 30\%. We chose to use Wafer 3 for our estimation because it has the largest amount of silicon upstream of  the targets, allowing for the closest compensation of the ejection through implantation. However, for \ber~there is also a contribution of implantation from production in the $\sim$\num{0.5}" of air between the wafer targets, which varies between \SIrange[range-phrase = --]{0.4}{0.6}{\milli\becquerel} for the different models. Because this is significant compared to the severely underestimated production and ejection in silicon for the BERTINI and BIC models, the ratio of the production to residual activity is also greatly underestimated and we have therefore chosen to not use the BERTINI and BIC models for estimations of the \ber~production rate from here onward. For all models without kinematic information we have used the ratio of production to residual \ber~activity from the Geant4 INCLXX model, i.e. \num{1.008 \pm 0.046}.

\subsubsection{\sod}
As seen in the third and fourth columns of Table~\ref{tab:sod_pred}, both the ejection and implantation fraction of \sod~nuclei are negligible due to the large size of the residual nucleus and no correction needs to be made to the predicted production activity.

%\todo{add figures of transmat}

\begin{figure}
\centering
\includegraphics[width=\columnwidth]{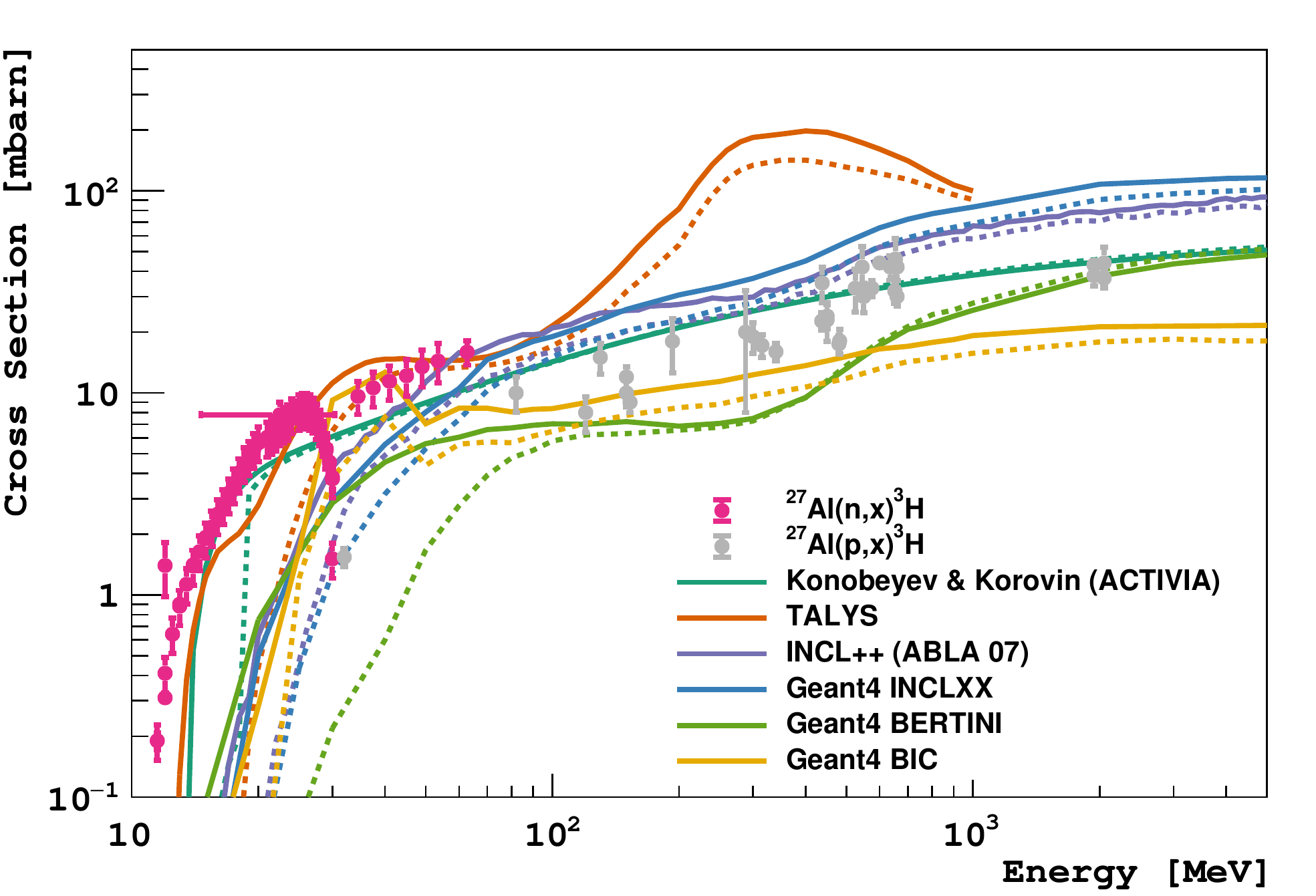}
\caption{Experimental measurements (data points) and model estimates (continuous lines) of the neutron-induced tritium production in aluminum. Measurements of the proton-induced cross section are also shown for reference. For direct comparison, we also show the corresponding model predictions for silicon (dashed lines) from Fig.~\ref{fig:si_3h_cross_sections}.}
\label{fig:al_3h_cross_sections}
\end{figure}

%\todo{Explain estimation of uncertainty}

\subsection{Comparison to Experimental Measurements}
The ratio of the experimentally measured activities to the predictions of the residual activity from different models are shown in the final column of Tables~\ref{tab:trit_pred}, \ref{tab:ber_pred}, and \ref{tab:sod_pred} for \trit{}, \ber{}, and \sod{} respectively. For tritium, it can be seen that the predictions of the K\&K and INCL models are in fairly good agreement with the measurement, while the TALYS model overpredicts and the Geant4 BERTINI and BIC models underpredict the activity by more than a factor of two.  For \ber, the best agreement with the data comes from the S\&T model and the spline fit to measurements of the proton-induced cross section. We note that the proton cross sections do slightly overpredict the production from neutrons, as found in Ref.~\cite{ninomiya2011cross}, but the value is within the measurement uncertainty. For \sod, there is good agreement between our measured activity and the predictions from the experimental measurements of the neutron-induced activity by Michel et al. \cite{michel2015excitation, hansmann2010production}, extrapolated at high energies using the TALYS model. For comparison, the use of the proton-induced production cross section (shown in Fig.~\ref{fig:si_22na_cross_sections}) leads to a value that is roughly 1.9$\times$ larger than our measured activity.

If we assume that the energy dependence of the cross-section model is correct, the ratio of the experimentally measured activity to the predicted activity is the normalization factor that must be applied to each model to match the experimental data. In the next section we will use this ratio to estimate the production rates from cosmic-ray neutrons at sea level.

\section{Cosmogenic Neutron Activation}
\label{sec:cosmogenic_rates}
The isotope production rate per unit target mass from the interaction of cosmic-ray neutrons, $P'$ [\si{\atoms\per\kg\per\second}], can be written as
\begin{linenomath*}
\begin{align}
P' = n \int \Phi(E) \cdot \sigma(E)~dE,
\end{align}
\end{linenomath*}
where $n$ is the number of target atoms per unit mass of silicon [atoms/kg], $\sigma(E)$ is the isotope production cross section [cm$^2$], $\Phi(E)$ is the cosmic-ray neutron flux [\si{\neutrons\per\cm\squared\per\second\per\MeV}], and the integral is evaluated from 1\,MeV to 10\,GeV.\footnote{The TALYS cross sections only extend up to 1 GeV \cite{koning2014extension}. We have assumed a constant extrapolation of the value at 1\,GeV for energies $>$1\,GeV.} While the cross section is not known across the entire energy range and each of the models predicts a different energy dependence, the overall normalization of each model is determined by the comparison to the measurements on the LANSCE neutron beam. The similar shapes of the LANSCE beam and the cosmic-ray neutron spectrum allow us to greatly reduce the systematic uncertainty arising from the unknown cross section.  

There have been several measurements and calculations of the cosmic-ray neutron flux (see, e.g., Refs.~\cite{hess1959cosmic, armstrong1973calculations, ziegler1996terrestrial}). The intensity of the neutron flux varies with altitude, location in the geomagnetic field, and solar magnetic activity---though the spectral shape does not vary as significantly---and correction factors must be applied to calculate the appropriate flux \cite{desilets2001scaling}. The most commonly used reference spectrum for sea-level cosmic-ray neutrons is the so-called ``Gordon'' spectrum \cite{gordon2004measurement} (shown in Fig.~\ref{fig:lanscebeamenergy}), which is based on measurements at five different sites in the United States, scaled to sea level at the location of New York City during the mid-point of solar modulation. We used the parameterization given in Ref.~\cite{gordon2004measurement}, which agrees with the data to within a few percent. The spectrum uncertainties at high energies are dominated by uncertainties in the spectrometer detector response function ($<4$\% below 10 MeV and 10--15\% above 150 MeV). We have assigned an average uncertainty of 12.5\% across the entire energy range.

\begin{table}[t!]
   \centering
   \begin{tabular}{ccc} 
   \hline
   Model & Predicted & Scaled \\
    & cosmogenic \trit & cosmogenic \trit \\
   & production rate & production rate\\
   & [\si{\atoms\per\kilogram\per\dayshort}] &  [\si{\atoms\per\kilogram\per\dayshort}] \\
   \hline
   K\&K (ACTIVIA) & \num{98 \pm 12} & \num{108 \pm 20} \\
   TALYS & \num{259 \pm 33} & \num{96 \pm 18}\\
   INCL++(ABLA07) & \num{106 \pm 13} & \num{114 \pm 22}\\
   G4 BERTINI & \num{36.1 \pm 4.5} & \num{124 \pm 22}\\
   G4 BIC & \num{42.8 \pm 5.4} & \num{102 \pm 17}\\
   G4 INCLXX & \num{110 \pm 14} & \num{130 \pm 23}\\
   \hline
   \end{tabular}
   \caption{Predicted \trit\ production rates (middle column) from sea-level cosmic-ray neutron interactions in silicon for different cross-section models. The final column provides our best estimate of the production rate for each model after scaling by the ratio of the measured to predicted \trit~activities for the LANSCE neutron beam.}
   \label{tab:trit_cosmic}
\end{table}

\begin{table}[t!]
   \centering
   \begin{tabular}{ccc} 
   \hline
   Model & Predicted & Scaled  \\
   & cosmogenic \ber & cosmogenic \ber \\
   & production rate & production rate\\
   & [\si{\atoms\per\kilogram\per\dayshort}] &  [\si{\atoms\per\kilogram\per\dayshort}] \\
   \hline
   S\&T (ACTIVIA) & \num{8.1 \pm 1.0} & \num{8.7 \pm 1.6}\\
   TALYS & \num{4.17\pm 0.52} & \num{6.2 \pm 1.3}\\
   INCL++(ABLA07) & \num{2.81 \pm 0.35} & \num{8.8 \pm 1.9}\\
   $^{\text{nat}}$Si(p,x)$^7$Be Spl. & \num{9.8 \pm 1.2} & \num{8.3 \pm 1.7}\\
   %GEANT4 BERTINI & \num{0.117 \pm 0.015} & \num{31.4 \pm 9.5} \\
   %GEANT4 BIC & \num{0.0394 \pm 0.0049} & \num{8.7 \pm 2.5}\\
   G4 INCLXX & \num{0.411 \pm 0.052} & \num{8.4 \pm 1.7}\\
   \hline
   \end{tabular}
   \caption{Predicted \ber\ production rates (middle column) from sea-level cosmic-ray neutron interactions in silicon for different cross-section models. The final column provides our best estimate of the production rate for each model after scaling by the ratio of the measured to predicted \ber~activities for the LANSCE neutron beam.}
   \label{tab:ber_cosmic}
\end{table}

\begin{table}[t!]
   \centering
   \begin{tabular}{ccc} 
   \hline
   Model & Predicted & Scaled  \\
    & cosmogenic \sod & cosmogenic \sod\\
   & production rate & production rate\\
   & [\si{\atoms\per\kilogram\per\dayshort}] &  [\si{\atoms\per\kilogram\per\dayshort}] \\
   \hline
   S\&T (ACTIVIA) & \num{86 \pm 11} & \num{43.2 \pm 7.1}\\
   TALYS & \num{60.5 \pm 7.6} &\num{43.0 \pm 6.8}\\
   INCL++(ALBA07) & \num{60.0 \pm 7.5} & \num{43.1 \pm 7.2}\\
   Michel-TALYS & \num{42.8 \pm 5.4} & \num{42.0 \pm 6.8}\\
   G4 BERTINI & \num{28.0 \pm 3.5} & \num{43.0 \pm 7.3}\\
   G4 BIC & \num{115 \pm 14} & \num{43.4 \pm 7.2}\\
   G4 INCLXX & \num{116 \pm 15} & \num{43.1 \pm 7.1}\\
   \hline
   \end{tabular}
   \caption{Predicted \sod\ production rates (middle column) from sea-level cosmic-ray neutron interactions in silicon for different cross-section models. The final column provides our best estimate of the production rate for each model after scaling by the ratio of the measured to predicted \sod~activities for the LANSCE neutron beam.}
   \label{tab:sod_cosmic}
\end{table}

The predicted production rates per unit target mass for the cross-section models considered are shown in the second columns of Tables~\ref{tab:trit_cosmic}, ~\ref{tab:ber_cosmic}, and~\ref{tab:sod_cosmic} for \trit, \ber, and \sod~respectively. Scaling these values by the ratio of the measured to predicted activities for the LANSCE neutron beam, we obtain our best estimates for the neutron-induced cosmogenic production rates per unit target mass, shown in the corresponding final columns. The spread in the values for the different cross-section models is an indication of the systematic uncertainty in the extrapolation from the LANSCE beam measurement to the cosmic-ray neutron spectrum. If the LANSCE neutron-beam spectral shape was the same as that of the cosmic-ray neutrons, or if the cross-section models all agreed in shape, the central values in the final column of each table would be identical. 

Our best estimate of the activation rate of tritium in silicon from cosmic-ray neutrons is \mbox{$(112 \pm 15_\text{exp} \pm 12_\text{cs} \pm 14_\text{nf})$} \si{\atomstrit\per\kg\per\day}, where the first uncertainty listed is due to experimental measurement uncertainties (represented by the average uncertainty on the ratio of the measured to predicted activities from the LANSCE beam irradiation for a specific cross-section model), the second is due to the uncertainty in the energy dependence of the cross section (calculated as the standard deviation of the scaled cosmogenic production rates of the different models), and the third is due to the uncertainty in the sea-level cosmic-ray neutron flux. Similarly, the neutron-induced cosmogenic activation rates for \ber\ and \sod\ in silicon are \mbox{$(8.1 \pm 1.3_\text{exp} \pm 1.1_\text{cs} \pm 1.0_\text{nf})$} \si{\atomsber\per\kg\per\day} and \mbox{$(43.0 \pm 4.7_\text{exp} \pm 0.4_\text{cs} \pm 5.4_\text{nf})$} \si{\atomssod\per\kg\per\day}.

\section{Activation from other particles}
\label{sec:alternate}
In addition to activity induced by fast neutrons, interactions of protons, gamma-rays, and muons also contribute to the total production rate of \trit, \ber~and \sod. In the following subsections we describe the methods we used to estimate the individual contributions using existing measurements and models. In some cases experimental data is very limited and we have had to rely on rough approximations based on other targets and related processes.

\subsection{Proton Induced Activity}
\begin{figure}
\centering
\includegraphics[width=\columnwidth]{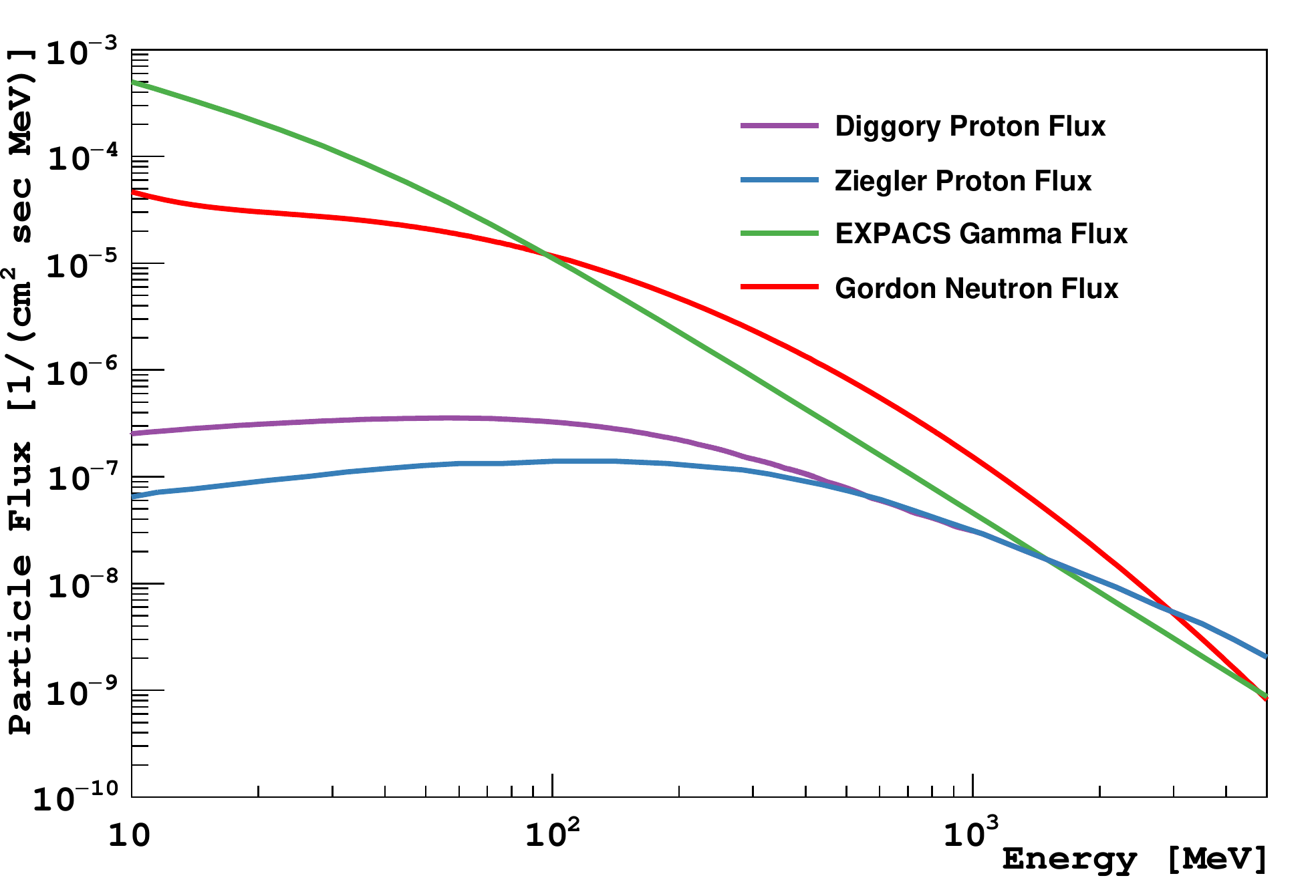}
\caption{Comparison of sea-level cosmic-ray fluxes of protons \cite{diggory1974momentum, ziegler1979effect, ziegler1981effect}, gamma rays \cite{expacs}, and neutrons \cite{gordon2004measurement}.}
\label{fig:alt_flux_comp}
\end{figure}

At sea level the flux of cosmic-ray protons is lower than that of cosmic-ray neutrons due to the attenuation effects of additional electromagnetic interactions in the atmosphere. To estimate the production rate from protons we have used the proton spectra from Ziegler \cite{ziegler1979effect, ziegler1981effect} and Diggory et.\ al.\ \cite{diggory1974momentum} (scaled by the angular distribution from the PARMA analytical model \cite{sato2016analytical} as implemented in the EXPACS software program \cite{expacs}), shown in Fig.~\ref{fig:alt_flux_comp}.

Experimental measurements of the proton-induced tritium production cross section have been made only at a few energies (see Fig.~\ref{fig:si_3h_cross_sections}). We have therefore based our estimates on the neutron cross-section models, scaled by the same factor used in Table~\ref{tab:trit_pred}. To account for possible differences between the proton- and neutron-induced cross sections, we have included a 30\% uncertainty based on the measured differences between the cross sections in aluminum (see Fig.~\ref{fig:al_3h_cross_sections}). Similar to the neutron-induced production, we have used the mean and sample standard deviation of the production rates calculated with all the different combinations of the proton spectra and cross-section models as our estimate of the central value and uncertainty, yielding a sea-level production rate from protons of \SI{10.0 \pm 4.5}{\atomstrit\per\kg\per\day}.

For \ber~and \sod, measurements of the proton cross section across the entire energy range have been made; we have used spline fits to the data with an overall uncertainty of roughly 10\% based on the experimental uncertainties (see Figs.~\ref{fig:si_7be_cross_sections}~and \ref{fig:si_22na_cross_sections}). Our best estimates for the \ber~and \sod~production rates from protons are \SI{1.14 \pm 0.14}{\atomsber\per\kg\per\day} and \SI{3.96 \pm 0.89}{\atomssod\per\kg\per\day}.

\begin{table*}[t!]
   \centering
   \begin{tabular}{cccc} 
   \hline
   \vrule width 0pt height 2.2ex
   Source & \trit~production rate & \ber~production rate & \sod~production rate \\
   & [\si{\atoms\per\kilogram\per\day}] & [\si{\atoms\per\kilogram\per\day}] &  [\si{\atoms\per\kilogram\per\day}] \\
   \hline
   Neutrons & \num{112 \pm 24} & \num{8.1 \pm 1.9} & \num{43.0 \pm 7.2}\\
   Protons & \num{10.0 \pm 4.5} & \num{1.14 \pm 0.14} & \num{3.96 \pm 0.89}\\
   Gamma Rays & \num{0.73 \pm 0.51} & \num{0.118 \pm 0.083} &  \num{2.2 \pm 1.5}\\
   Muon Capture & \num{1.57 \pm 0.92} & \num{0.09 \pm 0.09} & \num{0.48 \pm 0.11}\\
   \hline
   Total & \num{124 \pm 25} & \num{9.4 \pm 2.0} & \num{49.6 \pm 7.4}\\
   \hline
   \end{tabular}
   \caption{Final estimates of the radioisotope production rates in silicon exposed to cosmogenic particles at sea level.}
   \label{tab:final_cosmic_prod}
\end{table*}

\subsection{Gamma Ray Induced Activity}
\begin{figure}
\centering
\includegraphics[width=\columnwidth]{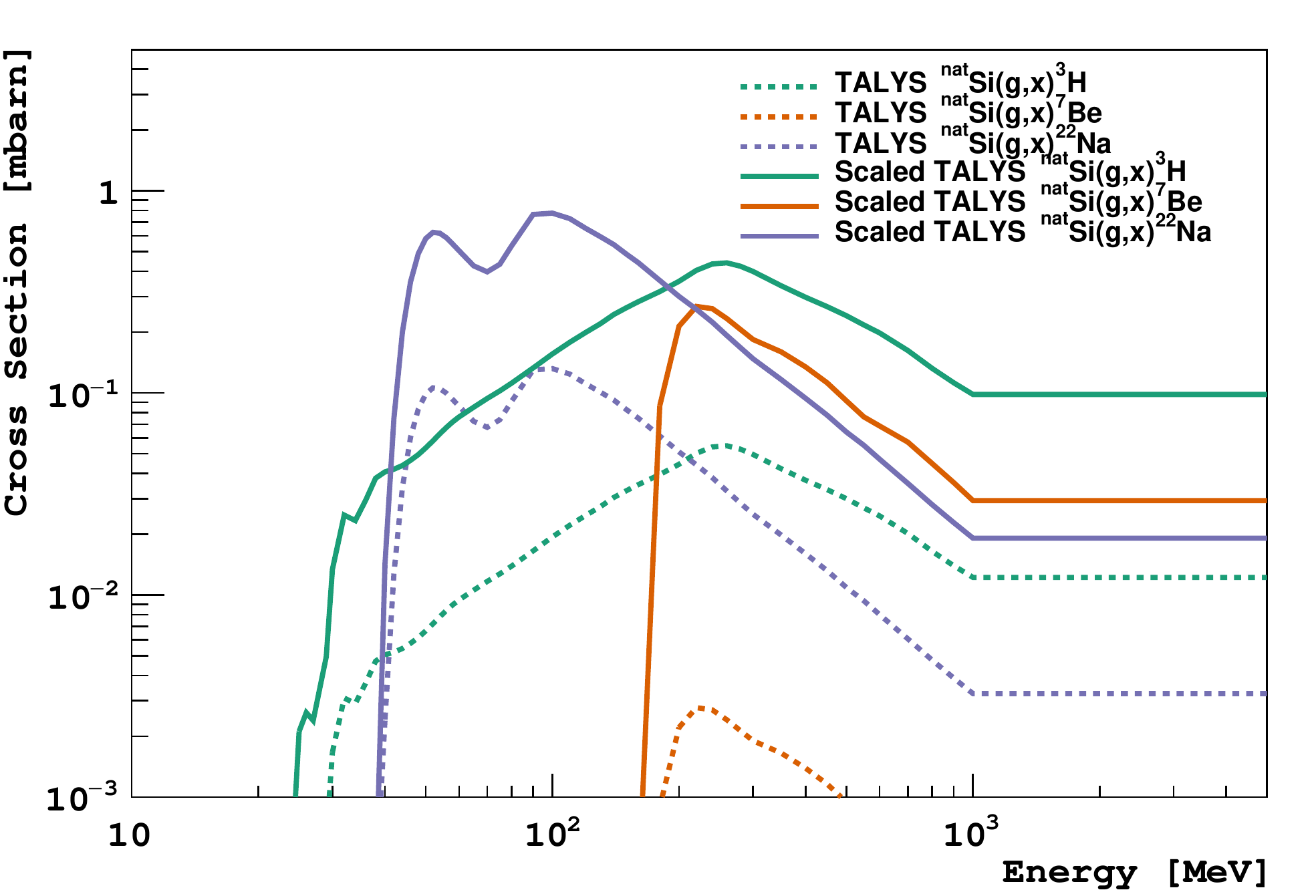}
\caption{Estimated photonuclear cross-section models for production of \trit, \ber, and \sod. The dashed lines indicate the original models from TALYS while the solid lines indicate the models scaled to match yield measurements made with bremsstrahlung radiation \cite{matsumura2000target, currie1970photonuclear}.}
\label{fig:gamma_cs}
\end{figure}

The flux of high-energy gamma rays at the Earth's surface was obtained using the PARMA analytical model \cite{sato2016analytical} as implemented in the EXPACS software program \cite{expacs}. Similar to the neutron spectrum, we used New York City as our reference location for the gamma spectrum, which is shown in Fig.~\ref{fig:alt_flux_comp}.

Photonuclear yields of \ber~and \sod~in silicon have been measured using bremsstrahlung beams with endpoints ($E_0$) up to \SI{1}{\giga\eV} \cite{matsumura2000target}. We are not aware of any measurements of photonuclear tritium production in silicon, though there is a measurement in aluminum with $E_0 =$ \SI{90}{\MeV} \cite{currie1970photonuclear} which we assume to be the same as for silicon. The yields, $Y(E_0)$, are typically quoted in terms of the cross section per equivalent quanta (eq.q), defined as
\begin{align}
    Y(E_0) = \frac{\displaystyle\int_0^{E_0} \sigma(k)N(E_0,k)dk}{\displaystyle \frac{1}{E_0}\int_0^{E_0} kN(E_0,k)dk}
\end{align}
where $\sigma(k)$ is the cross section as a function of photon energy $k$, and $N(E_0, k)$ is the bremsstrahlung energy spectrum.
To obtain an estimate for $\sigma(k)$, we assume a $1/k$ energy dependence for $N(E_0, k)$~\cite{tesch1971accuracy} and scale the TALYS photonuclear cross section models to match the measured yields of \SI{72}{\micro\barn \per \eqquanta} at $E_0 =$ \SI{90}{\MeV} for tritium and \SI{227}{\micro\barn \per \eqquanta} and \SI{992}{\micro\barn \per \eqquanta} at $E_0 =$ \SI{1000}{\MeV} for \ber\ and \sod , respectively (see Fig.~\ref{fig:gamma_cs}).
This corresponds to estimated photonuclear production rates of \SI{0.73}{\atomstrit\per\kilogram\per\day}, \SI{0.12}{\atomsber\per\kilogram\per\day}, and \SI{2.2}{\atomssod\per\kilogram\per\day}. Given the large uncertainties in the measured yields, the cross-section spectral shape, and the bremsstrahlung spectrum, we assume a $\sim 70\%$ overall uncertainty on these rates.

\subsection{Muon Capture Induced Activity}
%Muons stopped in silicon can be captured on a bound proton, releasing roughly \SI{100}{\MeV} of energy \cite{measday2001nuclear}. This energy is mainly taken away by the neutrino, but the nucleus can absorb significant energy, leading to several possible nuclear reactions. 
The production rate of a specific isotope $X$ from sea-level cosmogenic muon capture can be expressed as
\begin{align}
    P_\mu(X) = R_0 \cdot \frac{\lambda_c\text{(Si)}}{Q\lambda_d + \lambda_c\text{(Si)}}\cdot f_\text{Si}(X)
\end{align}
where $R_0 = \SI{484 \pm 52}{\muons\per\kg\per\day}$ is the rate of stopped negative muons at sea level at geomagnetic latitudes of about \SI{40}{\degree} \cite{charalambus1971nuclear}, the middle term is the fraction of muons that capture on silicon (as opposed to decaying) with the capture rate on silicon $\lambda_c$(Si) = \SI{8.712 \pm 0.018 E5}{\per\sec} \cite{suzuki1987total}, the decay rate of muons $\lambda_d$ = \SI{4.552E5}{\per\sec} \cite{tanabashi2018m}, and the Huff correction factor $Q = 0.992$ for bound-state decay \cite{measday2001nuclear}. The final term, $f_\text{Si}(X)$, is the fraction of muon captures on silicon that produce isotope $X$.

%The main mechanism for de-excitation following muon capture is neutron emission, though charged particles can also be emitted. 
For $^{28}$Si the fraction of muon captures with charged particles emitted has been measured to be \SI{15 \pm 2}{\percent} with theoretical estimates \cite{lifshitz1980nuclear} predicting the composition to be dominated by protons ($f_\text{Si}(^1$H) =  \SI{8.8}{\percent}), alphas ($f_\text{Si}(^4$He) = \SI{3.4}{\percent}), and deuterons ($f_\text{Si}(^2$H) = \SI{2.2}{\percent}). The total fraction of muon captures that produce tritons has not been experimentally measured\footnote{A direct measurement of triton production from muon capture in silicon was performed by the \href{http://muon.npl.washington.edu/exp/AlCap/index.html}{AlCap 
Collaboration} and a publication is in preparation. }, but a lower limit can be set at \SI{7 \pm 4 e-3}{\percent} from an experimental measurement of tritons emitted above 24 MeV \cite{budyashov1971charged}. %(compared to roughly \SI{0.32}{\percent} and \SI{0.15}{\percent} for protons and deuterons respectively above that same threshold). 
Recent measurements of the emission fractions of protons and deuterons following muon capture on aluminum have found values of $f_\text{Al}(^1$H) = \SI{4.5 \pm 0.3}{\percent} and $f_\text{Al}(^2$H) = \SI{1.8 \pm 0.2}{\percent} \cite{gaponenko2020charged}, and those same data can be used to calculate a rough triton emission fraction of $f_\text{Al}(^3$H) = \SI{0.4}{\percent} \cite{gaponenkopersonal}. If one assumes the same triton kinetic energy distribution in silicon as estimated for aluminum \cite{gaponenko2020charged} and uses it to scale the value measured above 24 MeV, one obtains a triton production estimate of $f_\text{Si}(^3$H) = \SI{0.49 \pm 0.28}{\percent}. The production rate of tritons from muon capture is then estimated to be \SI{1.57 \pm 0.92}{\atomstrit\per\kg\per\day}. 

The fraction of muon captures that produce \sod~has been measured at $f_\text{Si}$(\sod) = \SI{0.15 \pm 0.03}{\percent} \cite{heisinger2002production}, corresponding to a production rate from muon captures of \SI{0.48 \pm 0.11}{\atomssod\per\kg\per\day}. To our knowledge there have been no measurements of the production of \ber~through muon capture on silicon. We assume the ratio of \ber~to \sod~production is the same for muon capture as it is for the neutron production rates calculated earlier, with roughly \SI{100}{\percent} uncertainty, resulting in an estimated production rate from muon captures of \SI{0.09 \pm 0.09}{\atomsber\per\kg\per\day}.

\section{Discussion}
\label{sec:discussion}
The final estimates for the total cosmogenic production rates of \trit, \ber, and \sod~at sea level are listed in Table~\ref{tab:final_cosmic_prod}. These rates can be scaled by the known variations of particle flux with altitude or depth, location in the geomagnetic field, and solar activity, to obtain the total expected activity in silicon-based detectors for specific fabrication, transportation, and storage scenarios. The production rate at sea level is dominated by neutron-induced interactions, but for shallow underground locations muon capture may be the dominant production mechanism. For estimates of the tritium background, implantation of tritons generated in surrounding materials and ejection of tritons from thin silicon targets should also be taken into account.

Tritium is the main cosmogenic background of concern for silicon-based dark matter detectors. At low energies, 0--5\,keV, %\SIrange[range-phrase=--]{0}{5}{\keV}, 
the estimated production rate corresponds to an activity of roughly \SI{0.002} {\decays \per \keV \per \kg \per \day} per day of sea-level exposure. This places strong restrictions on the fabrication and transportation of silicon detectors for next-generation dark matter experiments. In order to mitigate the tritium background we are currently exploring the possibility of using low-temperature baking to remove implanted tritium from fabricated silicon devices.  

Aside from silicon-based dark matter detectors, silicon is also widely used in sensors and electronics for rare-event searches due to the widespread use of silicon in the semiconductor industry and the availability of high-purity silicon. The relative contributions of \trit, \ber, and \sod~to the overall background rate of an experiment depends not only on the activation rate but also on the location of these components within the detector and the specific energy region of interest. The cosmogenic production rates determined here can be used to calculate experiment-specific background contributions and shielding requirements for all silicon-based materials.

\section{Acknowledgements}
We are grateful to John Amsbaugh and Seth Ferrara for designing the beamline holders, Larry Rodriguez for assistance during the beam time, and Brian Glasgow and Allan Myers for help with the gamma counting. We would also like to thank Alan Robinson and Andrei Gaponenko for useful discussions on production mechanisms from other particles. This work was performed, in part, at the Los Alamos Neutron Science Center (LANSCE), a NNSA User Facility operated for the U.S.\ Department of Energy (DOE) by Los Alamos National Laboratory (Contract 89233218CNA000001) and we thank John O'Donnell for his assistance with the beam exposure and data acquisition. 
Pacific Northwest National Laboratory (PNNL) is operated by Battelle Memorial Institute for the U.S.\ Department of Energy (DOE) under Contract No.\ DE-AC05-76RL01830; 
the experimental approach was originally developed under the Nuclear-physics, Particle-physics, Astrophysics, and Cosmology (NPAC) Initiative, a Laboratory Directed Research and Development (LDRD) effort at PNNL, while the application to CCDs was performed under the DOE Office of High Energy Physics' Advanced Technology R\&D subprogram. We acknowledge the financial support from National Science Foundation through Grant No.\ NSF PHY-1806974 and from the Kavli Institute for Cosmological Physics at The University of Chicago through an endowment from the Kavli Foundation. The CCD development work was supported in part by the Director, Office of Science, of the U.S. Department of Energy under Contract No. DE-AC02-05CH11231.

\end{document}